\documentclass{elsart}
\usepackage{epsfig}
\usepackage{amssymb}
\journal{Physics Reports}

\newcommand{\msun}{\mbox{$M_\odot$}}

\def\be{\begin{eqnarray}}
\def\ee{\end{eqnarray}}
\def\lsim{\mathrel{\rlap{\lower3pt\hbox{\hskip1pt$\sim$}}
     \raise1pt\hbox{$<$}}} 
\def\gsim{\mathrel{\rlap{\lower3pt\hbox{\hskip1pt$\sim$}}
     \raise1pt\hbox{$>$}}} 
\def\la{\langle}
\def\ra{\rangle}

\def\cal{\it}

\begin{document}

\runauthor{Brown, \& Lee}

\begin{frontmatter}
\title{Recent Developments on Kaon Condensation and Its Astrophysical Implications}

\author[suny]{Gerald E. Brown,}
\author[pnu]{Chang-Hwan Lee,}
\author[saclay]{Mannque Rho}

\address[suny]{Department of Physics and Astronomy,
               State University of New York, Stony Brook, NY 11794, USA}

\address[pnu]{Department of Physics, Pusan National University,
              Busan 609-735, Korea} 

\address[saclay]{Institut de Physique Th\'eorique,
 CEA Saclay, 91191 Gif-sur-Yvette c\'edex, France}


\begin{abstract}
We discuss three different ways to arrive at kaon condensation at
$n_c\simeq 3 n_0$ where
$n_0$ is nuclear matter density: (1) Fluctuating around the $n=0$ vacuum
in chiral perturbation theory, (2) fluctuating around $n_{VM}$
near the chiral restoration density $n_\chi$
where the vector manifestation of hidden local symmetry is reached
and (3) fluctuating around the Fermi liquid fixed point at $\sim n_0$. They all share one common theoretical basis, ``hidden local symmetry."
We argue that when the critical density $n_c < n_\chi$ is reached in a
neutron star, the electrons turn into $K^-$ mesons, which go into an
S-wave Bose condensate. This reduces the pressure substantially and
the neutron star goes into a black hole.
Next we develop the argument that the collapse of a neutron star into a black hole takes
place for a star of $M \simeq 1.5\msun$. This means that Supernova 1987A
had a black hole as result. We also show that two neutron stars in a
binary have to be within 4\% of each other in mass, for neutron
stars sufficiently massive that they escape helium shell burning.
For those that are so light that they do have helium shell burning,
after a small correction for this they must be within 4\% of each
other in mass. Observations support the proximity in mass inside of
a neutron star binary.
The result of strangeness condensation is that there are $\sim 5$
times more low-mass black-hole, neutron-star binaries than double
neutron-star binaries although the former are difficult to observe.
\end{abstract}

\end{frontmatter}

\newpage

\tableofcontents

\newpage

\section{INTRODUCTION}
While the phase structure of hadronic matter at high temperature
both below and above the chiral phase transition temperature
$T_\chi$ is being mapped out by laboratory experiments with
invaluable help from lattice QCD, the situation with high density is
vastly different. There is little information about matter above the
nuclear matter density $n_0\approx 0.16$ fm$^{-3}$ from laboratory
experiments and there is practically no guidance from QCD proper
since lattice -- the only non-perturbative QCD tool available --
cannot handle high density and the reliable perturbative QCD
approach can access density regime only at asymptotic density where
such novel phenomena like color-flavor-locked superconductivity can
take place. To theorists' disappointments, though, this density
regime may be totally irrelevant to nature. We have a complicated
landscape of phases theoretically predicted near and above a chiral
restoration $n_\chi$ which is as rich as the phase structure of
water but they are all based on models, the reliability of which is
uncertain given the paucity of experimental supports and lack of
reliable theoretical control. Consequently the plethora of different
scenarios for dense stellar systems such as neutron stars, black
holes etc. available in the literature offer little guidance for
understanding compact star physics.

The aim of this paper is to exploit a systematic effective field
theory framework from three different vantage points to describe with some confidence the phase
structure of dense matter $crucially$ relevant to the formation of
stable compact stars and its implications on the population of
neutron stars and light-mass black holes.  These approaches, particularly the second and the third, rely on one common strategy that hadronic systems under extreme conditions can be accessed reliably by hidden local symmetry~\cite{HY:PR} that incorporates the scaling property of chiral symmetry~\cite{BR91}. Our reasoning will be
backed by detailed analysis of astrophysical observations.

This paper consists of two parts, one on an important phase change in hadronic physics
and the other on collapse to black holes in astrophysics.
Our aim is to bridge these seemingly disparate
branches of physical phenomena. In the first part, we discuss recent developments on the
most likely phase transition in hadronic matter, namely, kaon condensation, as density increases
beyond the nuclear matter density to $\sim 3\ n_0$. We will develop the thesis that this is the first -- and perhaps the last --
crucial phase change at high density that matters for the fate of compact stars, leaving wide-open the
possibility of other forms of higher-density phases involving quark matter etc. In
the second part, we give compelling arguments why astrophysical observations
strongly support that neutron stars of mass greater than $\sim
1.5 \msun$ (where $\msun$ is the solar mass) cannot be stable,
as a consequence of which any compact star more massive
than the maximum stable mass must be in the form of a black hole.
These two developments will then be joined to arrive at the conclusion that
kaon condensation at $\sim 3n_0$ implies $\sim 5$
times more low-mass black-hole, neutron-star binaries than double
neutron-star binaries.

\section{KAON CONDENSATION}
Although QCD cannot provide at present useful and quantitative
information for hadronic interactions at high density relevant to
the physics of compact stars, there are three vantage points at
which we have available reliable effective field theory tools to
work with. The first is the matter-free $T=n=0$ vacuum about which
fluctuations can be described by effective chiral field theory. Here
there is a wealth of experimental data to guide model building and
extrapolating beyond the normal matter density. The second is the
other extreme regime of high temperature and/or density at which
chiral symmetry is supposed to be restored, namely, the ``vector
manifestation fixed point" in hidden local symmetry formulated by
Harada and Yamawaki~\cite{HY:PR} characterized by the gauge coupling
going to zero and the interactions between hadrons becoming weak.
The third is in between the two limits, i.e., in the vicinity of
nuclear matter density at which the Fermi-liquid fixed point is
located. Here both theoretical and experimental information
accumulated since many decades in nuclear physics offers both
guidance and control near the nuclear matter density and indicates
what needs to be taken into account beyond. The results of these
three approaches giving the critical density $n_c\lsim 3 n_0$ have
been reported before. What is new is the conceptual link between
them provided by the recent development of chiral dynamics in hidden
local symmetry theory recently supported by a holographic dual
approach in string theory that gives an effective field theory of
QCD in terms of (infinite towers of both) vector mesons and baryons
and pions.

We will approach the state of matter believed to be present in the interior of
compact stars from these three vantage points and converge to the
result that kaons must condense in the vicinity of $3 n_0$. This then leads us
to propose that kaon condensation is the most likely phase
transition to take place in hadronic matter in the density regime
{\it that is relevant} to the fate of compact stars. Any other possible phases
such as e.g., quark matter, color superconductivity etc. that can appear at higher
densities will then be academic issues for stellar objects, although interesting
from a purely theoretical point of view. As we shall outline, the
metastable behavior of SN1987A which gave off neutrinos for $\sim
12$ seconds and then disappeared in all respects is symptomatic of a
condensate of negative charge kaons. In a number of papers we have
outlined how evidence of a condensate which sends neutron stars into
low-mass black holes, black holes not much more massive than neutron
stars, is already present, and to us quite convincing, in the
spectrum of masses in neutron-star binaries.

It seems highly likely that the densities $\sim 3 n_0$ are reached
in neutron stars. It is
touch or go whether they were reached in the progenitor of SN1987A.
However, they certainly are reached in the evolution of binary
neutron stars in the case where the two progenitor giant masses are
more than $4\%$ different from each other in which case the first
born neutron star finds itself in the red giant envelope of the
companion giant as the latter evolves. This neutron star accretes
matter from the evolving companion giant hypercritically until its
mass is increased $\sim 0.7\msun$, reaching densities enough for
kaon condensation which sends it into a low-mass black hole.

If, on the other hand, the two progenitor giants are within 4\% of
each other in mass, they will burn helium at the same time and avoid
the red giant stage of the second giant, which will then end up as a
neutron star.

\subsection{Kaon Condensation As Restoration Of
Explicit Chiral Symmetry Breaking} As the first estimate of the
critical density for kaon condensation, we approach dense hadronic
matter from the $T=n=0$ vacuum. This can be efficiently done by
resorting to Weinberg's ``folk theorem" \cite{weinberg} using
effective chiral field theory. The basic idea is to construct an EFT
Lagrangian that takes into account $all$ relevant degrees of freedom
at low energy and $all$ pertinent symmetry constraints, in
particular, chiral symmetry and do a systematic chiral perturbation
theory calculation to as high an order as feasible. This procedure
is by now fairly well established for pion-nucleon interactions as
well as, to some degree, for finite nuclei and is being currently
extended to nuclear matter~\cite{CND2}. The first approach made in
this spirit is the work by Kaplan and Nelson~\cite{kaplan-nelson}.
In this subsection, we describe a toy description that while perhaps oversimplified,
nonetheless brings out the essential feature of the physics involved while
ignoring what we consider to be inessential complications. The key
idea here is that kaon condensation can be viewed as the
$restoration$ by baryon density of chiral symmetry explicitly broken
by the strange quark mass.

To formulate this idea in the most economical way,
it was found to be simplest to take the V-spin projection of the Goldstone boson
sphere onto the V-spin circle, composed of $K^-$ and $\sigma$ in the $SU(3)$
chiral Lagrangian~\cite{V-spin}. The
Hamiltonian for explicit $K^-$ chiral symmetry breaking brings in
the kaon contribution to the explicit chiral symmetry breaking in
the nucleon $\Sigma_{KN}$ and in the kaon mass $m_K$. The
Hamiltonian for explicit chiral symmetry breaking -- which is of
${\cal O} (p^2)$ in the chiral power counting --
is
 \be H_{\chi SB}
& = & \Sigma_{KN}\langle\bar N N\rangle \cos\theta
+\frac 12 m_K^2 {f_\pi}^2 \sin^2\theta \nonumber\\
&\simeq & \Sigma_{KN} \langle \bar N N\rangle
\left(1-\frac{\theta^2}{2}\right) + \frac 12 m_K^2
{f_\pi}^2\theta^2
 \ee
where the last expression is obtained for small fluctuations
$\theta$. Here we have taken mean field in the baryon sector.
Dropping the term independent of $\theta$, we find
 \be
{m_K^\star}^2 = m_K^2 \left(1-\frac{\Sigma_{KN} \langle\bar N
N\rangle}{f_\pi^2 m_K^2}\right)^{1/2}. \label{eq2}
 \ee
The $\Sigma_{KN}$ is
 \be \Sigma_{KN} = \frac{(m_u+m_s) \langle
N|\bar u u + \bar s s|N\rangle} {(m_u+m_d) \langle N | \bar u u +
\bar d d| N\rangle} \sigma_{\pi N}
 \ee where $m_{u,d,s}$ are the
current quark masses and $\sigma_{\pi N}$ is the pion $\sigma$-term.
Lattice calculations \cite{Dong96} give
 \be \Sigma_{KN} \simeq 389(14) \; {\rm MeV}.
 \ee

We can go one order higher in the ``effective mass" of the kaon
(\ref{eq2}) by incorporating what is called ``range term" which
turns out be important quantitatively~\cite{Lee96}. It amounts to
changing $\Sigma_{KN}$ to
 \be \Sigma_{KN}^{\rm eff} =
\left(1-0.37\frac{{\omega_{K^-}^2}}{m_K^2}\right) \Sigma_{KN}
 \ee
where $\omega_{K^-}$ is the (anti)kaon energy
\be \omega_{K^-} =
V_{TW}^{K^-} + \sqrt{k^2+{m_{K^-}^\star}^2}.
 \ee

\begin{figure}
\centerline{\epsfig{file=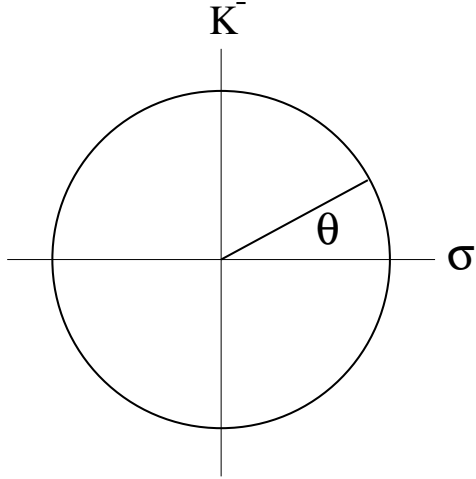,height=2.5in}}
\caption{Projection onto the $\sigma$-$K^-$ plane. The angular
variable $\theta$ represents fluctuation toward kaon mean field.}
\end{figure}

In addition to the scalar attraction (which can be thought of as
giving the kaon an effective mass), there is the vector attraction
given by the Tomozawa-Weinberg term \cite{Waas97} -- which is the
leading term, i.e., ${\cal O} (p)$, in the chiral counting
 \be V^{K^-}_{\rm
TW} = -\frac{3}{8 f_\pi^2} n \approx -60 \; {\rm MeV} \frac{n}{n_0}
\label{eq5}\ee where $n_0$ is nuclear matter density and we have
taken the empirical value for the pion decay constant
 \be f_\pi \simeq 93 \; {\rm MeV}.
\label{eq6}
 \ee

In a neutron star the electron chemical potential $\mu_{e^-}$, which
in the progenitor of SN1987A was $\sim 220$ MeV \cite{Thor94} is the
driving force towards kaon condensation~\cite{novel-mech}. Once the {\it in-medium}
kaon mass $m_{K^-}^\star$ is less than $\mu_e$ then the
degenerate electrons in the electron cloud in the neutron star can
lower the energy by changing into a kaon condensate. The kaons are
bosons, so they will immediately go into an S-state condensate,
which lowers the pressure substantially from that of the degenerate
electrons in the cloud in the neutron star. After this drop in
pressure, the neutron star will go into a black hole in a
light-crossing time.

The above captures the calculation of kaon condensation as it stood for many years
that indicated that kaons could condense at $n_c \lsim 3 n_0$. The principal lesson we learn here is that baryon density plays a prominent role, namely, that the underlying mechanism for condensation is the restoration of the explicitly broken chiral symmetry by baryon density. As noted in \cite{rge-lrs}, kaons condense because the s quark mass is neither too light nor too heavy.

The most complete calculation of possible kaon condensation along the
general line described above was
carried out by Thorsson et al. \cite{Thor94} with three different
$\Sigma_{KN}$, expressed in other terms in their chiral perturbation
theory. This calculation contained the medium dependence of the kaon
mass through $\Sigma_{KN}$ (see eq.~(\ref{eq2})) but did not contain
the medium dependence of $f_\pi$, the in-medium pion decay constant
as required by HLS theory which we believe is indispensable as we will see below.~\footnote{In HLS treatment, this is connected
to what is called ``intrinsic background dependence" resulting from
Wilsonian matching to QCD, the condensates of which are background --
temperature, density etc. -- dependent.}

One may raise objections to the above simple treatment on the ground
that several mechanisms that could be important are left out in the
treatment. Most obvious are the roles of the $\Lambda (1405)$ and
hyperons, much discussed in the literature, which when one does fluctuations
from the matter-free
vacuum, cannot be ignored. When looked at in terms of an RG flow, these
degrees of freedom give ``irrelevant" terms for the location of
the critical density, that is, insignificant for the precise value but they
play a determinant role in $triggering$ phase transitions.

Since the role of the $\Lambda (1405)$ in nuclei and dense matter
has been greatly disputed
in the past, we clarify what it as well as other related degrees of freedom
-- such as p-wave hyperon interactions -- that we will refer to as ``dangerously irrelevant operators" following condensed matter terminology,
do in the condensation phenomenon. The point we will drive at is
that while they may  play a crucial role in $driving$ the system toward
instability, once the direction in which the system moves is
determined, the location of the critical density is highly
insensitive to the strength of such interaction terms involved.

We focus on the contribution of $\Lambda (1405)$ to kaon-nuclear
interactions.~\footnote{Similar arguments can be made for p-wave hyperon
interactions. They give rise to ``irrelevant" four-Fermi interactions.}
The $\Lambda (1405)$-nucleon interactions can be
written as four-Fermi interactions which have canonical dimension
6. As such, it is ``irrelevant" from the usual dimensional counting. To
show what such a term does in meson condensation, we
follow the argument given in \cite{rge-lrs} for generic meson
condensation in strong interaction physics. It is a toy-model argument
but it is applicable to the present case. The technique used is
the one developed by Shankar~\cite{shankar} for Landau Fermi liquid
theory. There is a distinct difference, however, between boson
condensation and Fermi liquid in that while a ``mass" term is
$relevant$ in the renormalization group sense, so figures
importantly in the condensation process, the effective mass in the Fermi
liquid is a fixed-point quantity, so it does not flow. This is
because the Fermi surface in the latter is fixed by fiat.

Let the meson field be denoted generically as $\Phi$. We will focus on two
terms, one ``mass" term and the other interaction term, with the
action written schematically (apart from the kinetic energy term)
 \be
S=S_M + S_I
 \ee
where
 \be
S_M &=& \int d\omega d^3q \tilde{M} \Phi^*\Phi,\\
S_I &=& \int (d\omega d^3q)^2 (d\epsilon d^3k)^2 h
\Phi^*\Phi\Psi^\dagger\Psi\delta^4 (\omega, \epsilon, q, k)
 \ee
where $\Psi$ is the baryon field containing nucleons, hyperons and
$\Lambda (1405)$ etc. Of course both $\tilde{M}$~\footnote{Tilde
represents that it contains other ``mass-like" terms.} and $h$
subsume many terms with varying powers in the suitable expansion
(e.g., the chiral expansion in chiral perturbation
theory).~\footnote{For instance, $h$ will contain, aside from the
kaon mass, leading $O(p)$ term of $\sim
\frac{\omega}{f_\pi^2}K^\dagger K N^\dagger N$, the next-to-leading
order $O(p^2)$ terms, $\sim \frac{\Sigma}{f_\pi^2} K^\dagger K
N^\dagger N$ etc.} For our discussion we need not specify them here.

\begin{figure}[t]
\centerline{\psfig{file=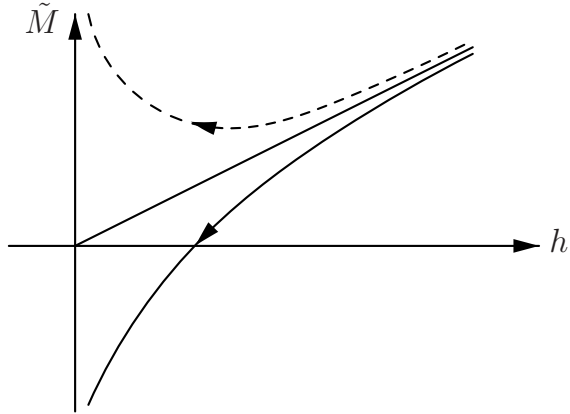,width=3.5in}} \caption{Schematic
diagram of RGE flow for the mass $\tilde{M}$. The dotted line with
arrow up shows the flow for $D <\frac 32 \frac{\tilde{M}_0}{h_0}$
and the solid line with arrow down for $D >\frac 32
\frac{\tilde{M}_0}{h_0}$.} \label{rg-lrs}
\end{figure}

The scaling law for the fields involved under scaling transformation
determined from the kinetic term (which is required to be invariant
under the scaling) shows that the ``mass" term is $relevant$ but the
interaction ($h$) term is $irrelevant$. Now the RGEs are found \`a
la Shankar~\cite{shankar} by
making the scale change $\Lambda\rightarrow s\Lambda$ with $0 < s <
1$ where $\Lambda$ is the cut-off scale~\cite{rge-lrs},
 \be
\frac{d\tilde{M}}{dt}&=& \tilde{M}-Dh,\\
\frac{dh}{dt}&=& -h/2
 \ee
where $t=-\ln s$ and $D=dn$ with $d>0$ is a constant depending on
$\Lambda/k_F$ where $k_F$ is the Fermi momentum of the baryonic
system. The solutions to the RGEs are
 \be
\tilde{M}(t)&=& (\tilde{M}_0 -\frac 23{Dh_0})e^t +\frac 23 Dh_0
e^{-t/2}, \\
h(t) &=& h_0 e^{-t/2}.
 \ee
As one decimates down with $s\rightarrow 0$ ($t\rightarrow \infty$)
integrating out higher lying modes, the boson-fermion interaction
gets exponentially suppressed. We see that if $D <\frac 32
\frac{\tilde{M}_0}{h_0}$, then the $\tilde{M}$ flows toward
$+\infty$. On the other hand, for
 \be
D >\frac 32 \frac{\tilde{M}_0}{h_0}\label{cdensity}
 \ee
the $\tilde{M}$ is bound to turn negative, signalling instability
against condensation. This can be seen in Figure \ref{rg-lrs}. Now
$D=dn$ and $d>0$, so that means that (\ref{cdensity}) can be
satisfied by dialling the density.  What matters is the ``least irrelevant term" figuring in $h_0$ that drives flow toward
condensation while more irrelevant terms -- unless of course numerically enhanced -- get suppressed.

Once the system, flowing toward instability, reaches the critical
point, then what matters will be the least ``irrelevant" term that
``eats up" the mass, with other irrelevant terms  being suppressed. In
fact a detailed analysis shows that the $\Lambda (1405)$ that brings the
initial attraction to direct the RG flow figures
negligibly in determining the location of the critical point~\cite{Lee96}:
it is found that change of the $\Lambda (1405)$-nucleon interaction
strength by orders of magnitude affects the critical density by only
a few \%.

Another objection that could be raised is that no account is made of
the strong repulsion that can be generated in nuclear interactions
in dense medium that would prevent kaon condensation at such a low
density. In ~\cite{pandharipandeetal}, Carlson et al. argue that
due account of short-range nuclear correlations could push the
critical density above $\sim 7 n_0$. This objection will be answered
in what follows in terms of ``hadronic freedom" in the vicinity of
the critical density which prevents the repulsion from growing at high density.
%

\subsection{Strangeness Condensation By Expanding About
the Fixed Point Of the Harada-Yamawaki Vector Manifestation} As the
second estimation of the critical density which we favor over the
others, we calculate kaon condensation going top-down from the
chiral restoration point. For this we adopt the hidden local
symmetry approach to chiral dynamics developed by Harada and
Yamawaki (HY)~\cite{HY:PR}. It is appealing to think of this theory
as a truncated infinite-tower description of holographic dual QCD
that arises from string theory (explained below) in which all vector
mesons lying above the lowest ones, $\rho$, $\omega$ and $\phi$, are
integrated out. This theory is characterized by that it has local
gauge invariance, which allows a systematic chiral perturbation
expansion with the vector mesons included and that the effective
theory is matched to QCD \`a la Wilsonian RGE and hence ``knows"
about QCD. The crucial outcome of the HY's analysis is that the
theory has what is called ``vector manifestation (VM)" fixed point
to which the theory flows subject to constraints imposed by QCD when
the system is driven by either temperature~\cite{HS:T-VM},
density~\cite{Hara02} or the number of flavors~\cite{HY:PR} to the
critical point at which chiral symmetry is restored. The fixed point
is given by
 \be
p^*:=(g^*, a^*, f_\pi^*)= (0, 1, 0)\label{VMfixedpoint}
 \ee
where $g$ is the hidden gauge coupling constant, $a=F_\sigma^2/F_\pi^2$ is
the ratio of ``parametric decay constants" of the pion $F_\pi$ and the
longitudinal component of the $\rho$ meson $F_\sigma$ and $f_\pi$ is the
physical pion decay constant.

What is highly pertinent to us is that near the VM fixed point,
fluctuations of the hadrons figuring in the theory become
well-defined, making processes taking place near
the VM fixed point amenable to a simple calculation. It has been
observed that even if a given process takes place away
from the VM fixed point, so $g\neq 0$ and $f_\pi\neq 0$, taking
$a\approx 1$ can be a good starting point for describing fluctuations in certain
processes. For example, $a=1$ is an accurate approximation for
describing the chiral doublers in heavy-light mesons such as $D$ and
$D^*$~\cite{chiraldouble}, the $\pi^+$-$\pi^0$ mass difference~\cite{HTY:pion},
the nucleon EM form factors~\cite{BR04} etc.

Assuming that kaons condense near, but below, the critical density for
chiral restoration $n_\chi$, we start from the VM fixed point
(\ref{VMfixedpoint}) and make a renormalization-group flow analysis as
density decreases from the VM fixed point as was done in
\cite{Brown06}. In performing this calculation, we exploit two
observations on: (1) the relevant degrees of freedom near the VM fixed point and
(2) the weakness of interaction strength in what is referred to as ``hadronic
freedom" zone.

We discuss these observations.

The first and most important observation is that the relevant
degrees of freedom for kaon condensation are not those manifest in
free space where the elementary kaon-nucleon interaction takes
place, but more relevant are the degrees of freedom in the vicinity
of the fixed point, since the condensation is in the neighborhood of
this point $n_{\chi}$ at which chiral restoration takes place. Thus,
for example, since $\langle\bar qq\rangle$ is near zero, the HLS
gauge coupling constant $g$ is near zero, making the vector mesons
nearly massless as pions are. This makes chiral perturbation
calculation feasible and reliable with the vector mesons and pions
treated on the same footing with their masses appearing as small
parameters of ${\cal O} (p^2)$. Furthermore, the $\Lambda (1405)$
and other excitations which are important near the matter-free
vacuum, become irrelevant in the sense defined above, so do not
figure in the calculation of the critical density.

We thus have vector mesons and pions as the explicit degrees of freedom.
But then what about baryons?

In holographic QCD which generalizes Harada-Yamawaki's HLS theory,
there are no fermion degrees of freedom. Baryons therefore must
arise as solitons. Indeed in the holographic dual QCD model of Sakai
and Sugimoto~\cite{Sakai05} which has correct features of chiral
symmetry properties of QCD proper, baryons do arise naturally as
instantons in five dimensions in which low-energy QCD is dualized or
alternatively as skyrmions in four dimensions embedded in an
infinite tower of vector mesons. The chiral properties of the
resulting baryons, in particular, nucleons, are very well described
in the approximations adopted by the model~\cite{HRYY}. In order to
exploit the VM fixed point, we need to integrate out all except the
lowest members of the tower and treat hadonic matter in terms of
skyrmions emerging from HY's HLS theory. Indeed hadronic matter at
high density has been described in terms of skyrmion matter with an
infinite winding number~\cite{skyrme-lattice}. However there is an
indication that skyrmions are not stable at some density $n_{flash}$
above $n_0$~\cite{alkoferetal}. We will argue that this implies that
the nucleon goes over to constituent (or quasi) quarks at $n_{flash}
< n_\chi$. We will first develop the argument in temperature and
then apply it to the density case.

\subsection{Hadronic Freedom, the Flash Point And Why the Constituent
Quark Model Is Relevant} What happens when hadronic system is heated
to near the critical temperature $T_\chi$, i.e., from 120 to 175
MeV, was studied by Brown et al \cite{Brown06b} using the STAR
peripheral data \cite{STAR}. For the peripheral experiments 120 MeV
was the freezeout temperature; i.e., the hadrons leaving the system
at that temperature did not interact further. The upper temperature
$T=175$ MeV was taken to be the phase transition temperature
$T_\chi$ at which the VM dictates that the HLS coupling $g$ go to
zero and the parameter $a$ go to 1.
%
%
%
Brown et al. \cite{Brown06b} showed that the widths $\Gamma
(\rho\rightarrow 2\pi)$ for $\rho$ decay into two pions were zero at
$T=T_\chi$ and remained small until just before freezeout at $T=120$
MeV. This was because the $\Gamma$'s went as
$(m_\rho^\star/m_\rho)^5$ and the mass of the $\rho$ was taken to
increase from zero at $T_\chi$ to 90\% of the on-shell $\Gamma$ at 120
MeV. The latter value was called the ``flash point" because the
interactions of the $\rho$'s at that temperature was that of
(nearly) on-shell hadrons. The $\rho$-mesons in the experiment were
recreated by following the pion tracks in the time projection
chamber back to the vertex, identified as the $\rho$, for those
which were emitted in $\rho$-decay. The 120 MeV was thus not only
the freezeout temperature, but also the temperature at which the
hadrons turned into constituent quarks.

We learn from the above observation that the region of temperature
from $\sim 120$ to $T_{\chi}$ thus delineates a region in which
interactions are negligible, a region governed by the vanishing
coupling constant $g\sim \la\bar{q}q\ra\rightarrow 0$. We can also
deduce what happens to the $a$ parameter which is defined by
 \be
m_\rho^2=a F_\pi^2 g^2.
 \ee
As is known by the KSRF relation,
$m_\rho^2=2 F_\pi^2 g^2$, and also from the vector dominance of the
pion EM form factor, $a=2$ in matter-free space. In the large $N_c$
limit in which the bare HLS Lagrangian is defined by Wilsonian
matching, one finds $a\simeq 4/3$ with the $a=1$ point lying on the
RG trajectory matched to QCD. The $a=2$ point is not on that
trajectory, implying that in HLS theory, $a=2$ can only be an
accident. Indeed as mentioned above, nature prefers $a\approx 1$ at
least in the leading order of chiral perturbation theory. It has
been shown by Harada and Sasaki~\cite{HS:T} that in heat bath, the
vector dominance given by $a=2$ is strongly violated as $a$ goes to
1 when temperature goes toward $T_\chi$. This suggests that we
associate $a_{flash}\sim 4/3$ as the flash point at which
$T=T_{flash}\sim 120$ MeV. We interpret the region between $a \sim
4/3$ at 120 MeV and $a=1$ at $T_{\chi}$ to be a region in which the
relevant fermionic degrees of freedom are constituent quarks, with
the off-shell quarks flowing freely from higher to lower
temperature. Nothing detectable goes on here in this region -- it is
just a region of space-time without anything except off-shell
particles within it, hence called ``hadronic freedom" region.

In dense matter, the situation is much less clear, so we are going to
make a certain number of guesses and develop a scenario that
parallels the temperature scenario.~\footnote{There is an indication from a skyrmion description of dense matter that $a\approx 1$ at the ``pseudo-gap density" which is identified as the flash density~\cite{MR-halfskyrmion}. This may be a better approximation than $a\simeq 4/3$ adopted below.}

As mentioned, it is established that the vector manifestation fixed
point (\ref{VMfixedpoint}) is reached when density is at the
critical density $n_\chi$ at which $\la\bar{q}q\ra=0$~\cite{Hara02}. It
has also been verified that $m_\rho\rightarrow g\propto
\la\bar{q}q\ra\rightarrow 0$ in the vicinity of the critical density and
$(a-1)\propto (\la\bar{q}q\ra)^2\rightarrow 0$. However the question
is: where is the density flash point above which the system goes
into  the hadronic freedom region?  In order to answer this question
within the framework of HLS theory, we need the baryon degrees of
freedom and at present, there is no systematic calculation in HLS
theory that includes baryons. What we know now is described in
\cite{BR04}: Up to the nuclear matter density $n_0$, the available phenomenology
in nuclear processes indicates that the HLS coupling $g$ stays more
or less unchanged. Beyond $n_0$, we have no information from data but
our basic assumption is that the flash point $n_{flash}$ lies at the point from
which $g$ drops rapidly arriving at $g=0$ at $n=n_c$. From what we
discussed above, we deduce that it is at $n_{flash}$ that going
top-down, the constituent quarks go on-shell forming baryons as the
coupling constant becomes strong leaving the hadronic freedom
region. Now if one extrapolates linearly from $a=1$ at $n_\chi$
to $a=2$ at $n=0$, we arrive at $a\simeq 4/3$ when
the density reaches the density
 \be
n_{flash}\simeq 3 n_0.\label{nflash}
 \ee
This is the density counterpart of $T_{flash}\simeq 120$ MeV. The
hadronic freedom region is identified as the interval
$n_{flash}\lsim n_{HF}\lsim n_\chi$. The putative repulsion one
expects naively from the vantage point of the matter-free vacuum
cannot survive within this region and would not provide the
mechanism often invoked for stabilizing a neutron star at as large
as $\gsim 2\ \msun$~\cite{akmaletal}.\footnote{We can understand
this in terms of an effective chiral Lagrangian that incorporates
Brown-Rho scaling (or ``intrinsic density dependence (IDD)" in HLS
theory)~\cite{CND2}. Consider four-Fermi (baryon) interaction terms
in the effective Lagrangian relevant to the vector-meson channels:
 \be
\sum_i C_i^\star (\bar{N}\Gamma_i N)^2\approx
-\frac{{C_{\tilde{\omega}}^\star}^2}{2} (\bar{N}\gamma_\mu N)^2
-\frac{{C_{\tilde{\rho}}^\star}^2}{2}(\bar{N}\gamma_\mu\tau N)^2
+\cdots.\label{4-fermi-dense}
 \ee
Here the asterisk denotes the IDD (intrinsic density dependence) in
the parameters, {\it i.e.}, Brown-Rho scaling, appearing in the EFT
Lagrangian and the ellipsis stands for other terms that are allowed
by chiral symmetry. Suppose now that the parameters in the matrix
elements of the four-Fermi interactions (\ref{4-fermi-dense}) are
expanded in power of the density operator $\hat{n}=\bar{N}\gamma_0
N$. The resulting Lagrangian will contain six- and higher-Fermi
fields and when mean field is taken, will give rise to density
dependent parameters in the theory. This means that certain chirally
symmetric $n$-body (with $n > 2$) interactions are subsumed in the
IDD of the coefficients in (\ref{4-fermi-dense}) when truncated to
the four-Fermi terms. Since the mean field using the Lagrangian is
equivalent to doing Landau Fermi liquid theory~\cite{CND2,BHLR07},
certain important many-body effects are therefore encoded chirally
symmetrically in the IDD of the parameters.

In this formulation, the repulsion resides in the $\omega$ channel
in (\ref{4-fermi-dense}).  In Harada-Yamawaki  HLS
theory, the parametric mass of the $\rho$ and $\omega$ (or the
lowest member of the infinite tower in holographic QCD) drops as density increases whereas the
vector coupling $g$ remains more or less unchanged up to $n \sim
n_0$ and then decreases after the ``flash density" $n_{flash}\sim
2n_0$. This means the coefficient ${C_{\tilde{\omega}}^\star}^2$
will increase as $g^2/{m_\omega^*}^2$ with the falling mass up to,
say, $n\sim n_0$ or slightly above. This repulsion up to that
density provides the mechanism for the saturation of nuclear matter.
From $n_{flash}$, the VM (vector manifestation) starts becoming
operative, with the gauge coupling and the mass falling at the same
rate so that the ratio ${g^*}^2/{m_\omega^*}^2$ will remain constant
at higher density. The repulsion will cease accordingly.}

\subsection{Calculation Of the Critical Density $n_c$ For Kaon Condensation}

We now have all the pertinent degrees of freedom in the vicinity of
$n_\chi$. They are the constituent (or quasi) quarks, the (pseudo)Goldstone
bosons $\pi$ and $K$ and the vector mesons $\rho$ and $\omega$ with
their masses comparable to the pion mass. To describe kaon
condensation, we need to take into consideration interactions between
the kaon and the quasiquarks. The Lagrangian we have is HLS
Lagrangian that is matched to QCD at a density $n\lsim n_\chi$ at a
matching scale $\Lambda_M^*$. Now we need to calculate RGE flow in
theory with the gauge coupling $g$ and the parameter $a$ as
described above in the density region $n_{flash}\lsim n\lsim n_\chi$.
The only other quantity that we have to consider is the parametric
pion decay constant $F_\pi^*$, since kaons will couple to the quarks
with the constant $1/F_\pi^*$. We need this constant at the matching
scale $\Lambda_M^*$. For this we observe that $F_\pi^*$ does not
scale appreciably in the matter-free space as found by Harada and
Yamawaki~\cite{HY:PR}. We shall give an argument why we expect this to be similar in medium. For this, it will suffice to
to know what it is near the nuclear matter density. This information comes from the analysis of deeply bound pionic atoms
\cite{Suzu04}\footnote{A detailed review of the pionic atom data and
the evidence for $F_\pi^\star=0.8 F_\pi$ is given by E. Friedman and
A. Gal \cite{Frie07}.} from which we learned that the in medium parametric
$F_\pi$ must be decreased
\be F_\pi \rightarrow F_\pi^* (n_0)\approx f_\pi^\star (n_0)\approx
0.8 F_\pi \label{Fpi-ratio}
 \ee
$\sim 20\%$ at $n=n_0$. We have no systematic calculation of $F_\pi^*$ beyond the nuclear matter density. This is because $F_\pi^*$ runs with the {\em intrinsic density dependence} dictated by the Wilsonian matching of the HLS correlators to those of QCD, which is unknown since lattice techniques cannot handle density effects. However one can argue that it will stay more or less unchanged up to the critical density. In both temperature and density, when one goes beyond the flash point, the vector meson mass $m_V^*$ and the gauge coupling constant $g^*$ scale in the same way, i.e. proportionally to $\la\bar{q}q\ra^*$. Hence the ratio $g^*/m_V^*\sim 1/F_\pi^*$ -- with $a^*\approx 1$ -- will stay  more or less unchanged. This feature was exploited in the high-temperature case in \cite{HS:dilepton}. As for the density relevant to kaon condensation, we need not go all the way to the chiral restoration point but one can make a rough estimate of the ratio $F_\pi^* (n_\chi)/F_\pi (0)$ in a HLS model with quasiquarks~\cite{Hara02},
\be
F_\pi^* (n_\chi)/F_\pi\sim \sqrt{3/5}\approx 0.77.
\ee
It thus seems justified to take the scaling (\ref{Fpi-ratio}) operative from the flash point to the kaon condensation point. It was shown in \cite{Brown06} that the
Walecka vector mean field used so much in nuclear physics already had this increase in it, so that it is part of the usual phenomenology of nuclear physics.

Since the gauge coupling constant $g$ is near zero in the hadronic freedom
region, quantum loop corrections can be ignored. It suffices therefore to do
the mean field calculation (i.e., tree contributions). For the neutron
fraction $x_n$ and proton fraction $x_p$ in a compact star, the
potential felt by the kaon will then be
 \be
V_{K^-}= -\frac{1}{a^* {F_\pi^*}^2}
\left(\frac{x_n}{2}+ x_p\right) n.\label{Kpot}
 \ee
We have included $\rho$ as well as $\omega$ exchange so as to make a
typical neutron star charge up at density $n_{\chi}$. Note that
(\ref{Kpot}) is $not$ in linear-density approximation. In this
picture, this is the only important contribution to the effective
mass of the kaon. Other terms are suppressed by the hadronic freedom
effect. Now let us see how the effective kaon mass behaves near the
chiral phase transition density $n_\chi\sim 4 n_0$. For this we take
$a^*=1$ and $F_\pi^*/F_\pi\simeq 0.8$. Then we find
 \be
\frac{[{g_V^\star}^2/{m_V^\star}^2]_{\rm
fixed\ point}}{[g_V^2/m_V^2]_{\rm zero\ density}} = \frac{[a
F_\pi^2]_{\rm zero\ density}}{[a^\star {F_\pi^\star}^2]_{\rm fixed\
point}} = \frac{2}{0.8^2} \simeq 3.1.
 \ee
For $x_n\approx 0.9$ corresponding to the neutron-star matter, we
have from (\ref{Kpot})
 \be
V_{K^-} (n=4 n_0) \approx - 516\ {\rm MeV}.
 \ee
This suggests that the kaon must have a vanishing mass at the chiral
transition density.

Let us move down to the flash point $\sim 3 n_0$. At this density,
taking into account the flow of $a$, we find
 \be
m_K^*(\frac 34 n_\chi)\frac 43\simeq m_{K^-}/4 \simeq 165\ {\rm
}MeV.
 \ee
This is somewhat smaller but not too far from the electron chemical
potential expected at that density, $\sim 220$ MeV~\cite{Thor94}.
Considering the roughness of the estimate, it seems reasonable to
identify the flash point as the kaon condensation critical density
$n_c$. We note that were it not for the presence of the electron
Fermi sea, the flow of the star matter would end up at the
VM fixed point at $n_{\chi}$. However, the decay of electrons into kaons
stops the hidden local symmetry (HLS) flow.

\subsection{Kaon Condensation From the Fermi-Liquid Fixed
Point} The third way to arrive at the kaon condensation critical
density $n_c\sim 3 n_0$ is to fluctuate from the density at which
nuclear matter is saturated, namely, $n_0\approx 0.16$ fm$^{-3}$.
The basic idea has been presented in previous review articles by the
authors, the most recent of which is given in \cite{BHLR07}. As
discussed there, going above the nuclear matter density consists of
first writing an effective Lagrangian that describes nuclear matter
as the Fermi-liquid fixed point \`a la Shankar when the mean field
approximation is taken. Given such a Lagrangian with the parameters
determined with the ``intrinsic density dependence" as dictated by
the Wilsonian matching to QCD prescribed in HLS theory, one can make
a fluctuation around the Fermi liquid fixed point to go to higher
density. The effective Lagrangian which is anchored on chiral
symmetry is equivalent to Walecka's linear mean field Lagrangian
with the intrinsic density dependence (i.e. Brown-Rho scaling)
suitably taken into account.

To proceed to kaon condensation, we assume as above that the
relevant fermionic degrees of freedom are the constituent quarks.
We consider the constituent quark picture applicable at nuclear
matter density and above and that the nucleon is made of loosely
bound three quasiquarks and the kaon of loosely bound strange quark
and chiral anti-quasiquark. Now the mean field nuclear potential has two
parts, the vector potential $V_N$ and the scalar potential $S_N$,
which can be viewed as due to exchange of an $\omega$ meson and a
scalar meson (denoted here $s$),
 \be
V_N &=&\frac{9}{8{F_\pi^*}^2} n,\nonumber\\
S_N &=& -\frac{{g_s^*}^2}{{m_s^*}^2} n.
 \ee
Here we are introducing a fictitious scalar in the line of Walecka model
with scaling mass and coupling constant  but one can actually avoid the
scalar degree of freedom by writing the effective Lagrangian in
terms of a four-Fermi interaction (see, e.g., \cite{BHLR07}). Now
from the phenomenology of Walecka model, we know that
 \be
S_N (n_0)- V_N (n_0)\lsim -600\ {\rm MeV}.
 \ee
Taking into account of the fact that the kaon contains only one
non-strange quark compared with three in the nucleon, we expect the
kaon to feel the attractive potential
 \be
S_{K^-} + V_{K^-} \lsim -200\ {\rm MeV}.
 \ee
Now applying the same reasoning to the compact star matter with a
90\% neutron fraction, we find at $n\sim 2n_0$ that
 \be
\omega_{K^-} (2n_0) \sim 160\ {\rm MeV}.
 \ee
Since the electron chemical potential expected at this density is
173 MeV~\cite{Thor94}, we see that kaon condensation is expected to
take place already at $2n_0$ in this scenario. Thus this picture
predicts $n_c < 3n_0$ consistent with the two other estimates.

\subsection{Delayed Collapse Of Neutron Star}

It is not generally realized that the metastability of SN1987A, {\it
i.e.} the fact that it emitted neutrinos for about 12 seconds before
disappearing (and was ``never heard from again"), was probably aided
substantially by the presence of strange particles in the core
\cite{Prak95}. For this, the fact that the particles were strange
was not the important point; the strange particles with strange
quarks, be they the $K^-$ or the hyperon $\Sigma^-$, have negative
charges and additional protons would be needed to neutralize their
charge.
%
%
%
The additional protons would interact strongly attractively with the neutron main component in the neutron star and increase the binding energy of the star.
As the star collapses to a neutron
star, once the densities are $\sim 10^{12}$ g cm$^{-3}$ the
neutrinos are trapped. That is, they may try to random walk their
way out, in the frame of the infalling matter, but once the density
is $\sim 10^{12}$ g cm$^{-3}$ the mean free path for neutrino
scattering off the nucleons is only a few centimeters, so the
neutrino drift (random walk) velocity outwards becomes smaller than
the velocity of infalling matter. The neutrinos, above the trapping
density, equilibrate with the neutrons and protons so that
 \be
\mu_n= \mu_p + \mu_e - \mu_\nu, \ee and the neutrinos fill a Fermi
sea of Fermi energy $\sim 100$ MeV. Thus, the neutrinos contribute
to the pressure.

It has been considered likely that the neutron star has $\Sigma^-$
hyperons in it. Although the mass of the $\Sigma^-$ hyperon is 1197
MeV, much heavier than the neutron 940 MeV, the neutrons will be
degenerate. Since the admixture of $\Sigma^-$ will have large
negative charge, its presence would bring down the electron chemical
potential $\mu_e$ considerably \cite{Glen01}. The $\Sigma^-$ at rest
can replace both an electron and a neutron. It begins to come in at
a density such that
 \be
\mu_{\Sigma^-}=\mu_n + \mu_e.
 \ee
This is generally satisfied at a density $n\sim 2 n_0$; i.e., at
about twice nuclear matter density \cite{Brow98c}. (The $\Lambda$'s
come in earlier, at a lower density, but the $\Lambda$ is charge
neutral and does not have the interesting effects that the
$\Sigma^-$ has.) As the neutrinos leave, the electrons recombine
with the protons to make neutrons and $\Sigma^-$'s, and the EOS
stiffens as the number of protons decreases. Thus, the star will not
collapse immediately into a black hole. Kaon condensation will set
in later as the star cools down further.

As has been discussed extensively in the literature,  the appearance
of hyperons in dense matter has been invoked to argue against kaon
condensation setting in at a density $\lsim 3n_0$ because the
driving force, {\it i.e.}, the electron chemical potential, would
get weakened~\cite{Glen01}. We believe this argument is not valid. We have made
clear that nucleons and hyperons are no longer the relevant degrees
of freedom when the flash point, which we estimate to be at $\sim 3
n_0$ is reached.
The RG flow is run completely by the Weinberg-Tomozawa vector interaction here, the hyperons having been integrated out, as irrelevant. Our fluctuation about the fixed point rather than about the perturbative vacuum has given a scaling which increases the vector interaction by $(a^\star)^{-1}$, increasing from 0.5 in free space to 1 at the fixed point. This explains much of the increased attraction and, therefore, softening of the EOS in our nonperturbative calculation.
%

\section{A SCENARIO FOR A LARGE NUMBER OF LOW-MASS BLACK HOLES IN THE GALAXY}

This was the title of a paper written by G.E. Brown and H.A. Bethe
in W.K. Kellogg Radiation Laboratory \cite{Brown94}. Since that
time, much improvements on the calculations have been made, the most
important one of which being the expanding about the VM fixed point
of hidden local symmetry. And furthermore more observational data
have been accumulated.

In this second half of the paper, we would like to present a chain
of astrophysical observations to link the kaon condensation in
hadronic matter developed in the first half to the population of
neutron stars and black holes. The most pertinent and daring
prediction on this issue had been made by Brown \cite{Brown95}, who
had found the accepted scenario for binary neutron star evolution to
be completely wrong. As we shall outline, the standard scenario
accepted up to then assumed that the accretion onto the first-born
neutron star, while the companion giant evolved and the neutron star
and giant went into common envelope, was limited by the Eddington
rate and, therefore, negligible. It was noted in \cite{Brown94},
after the suggestion by Chevalier \cite{Chev93}, that the rate of
accretion necessitated going over to hypercritical accretion and
that that was sufficient to send the first-born neutron star into a
(low-mass) black hole. It was then suggested that the only way to
avoid this was for the two progenitor giants to burn helium at the
same time, which allowed the neutron star to avoid the red giant
stage of the companion giant. This point is further developed below.

The Brown scenario had the prediction that the two neutron stars in
a binary, which must have burned helium at the same time, should be
within 4\% of each other in mass in order to do this. As we shall
see, this prediction is fulfilled by the most massive neutron star
binaries.

In the case of the less massive double neutron stars, {\it i.e.},
J0737-3039 and J1756$-$2251, the pulsars acquire an additional
$0.1-0.2\msun$ during the helium red giant stage of the companion,
which does not take place in the more massive binaries, but aside
from this, the pulsar and companion must be within 4\% of each other
in mass.
The companion in the less massive double neutron stars, i.e., J0737-3039 and J1756-2251 have to burn hotter than those in the more massive ones during the helium burning in order to reach the same central temperature of the more massive ones because the surface to volume ratio is larger in the less massive stars. This means that more energy is lost from the surface. The result of this is that the progenitors of these two less massive stars go through a helium red giant stage as well as (earlier) a hydrogen red giant stage. During the He red giant stage the pulsar accretes $0.1-0.2\msun$ of helium. Therefore, the necessary limit of at most 4\% difference between progenitor masses of the pulsar and companion can be somewhat larger if the companion of the pulsar gives an additional 0.1 to $0.2\msun$ of helium during the helium shell burning. Note that the helium burning of pulsar and companion progenitors must overlap in time. The helium red giant stage is put into the evolution of J0737-3039 by Willems and Kalogera \cite{Wil03}.

We will try to make our arguments for the general physics community.
The general reader would benefit from references to a few review papers, where neutron star structure \cite{Lat01,Web05,Lat06}, supernova explosions \cite{Jan01,Bur07,Jan07} and binary evolution \cite{Mez05,Pos06} are discussed.

\subsection{Hypercritical Accretion Onto First-Born Neutron Star
During Common Envelope Evolution}

In the study of neutron stars, a key tool used in astrophysics is
the acceptance of Eddington accretion as the maximum possible rate.
Matter falling on a neutron star is bound to the neutron star
gravitationally by $\sim 20\%$ of its rest mass because of the great
mass of the neutron star. This binding energy is used to produce
radiation, say in the form of photons, which are emitted from the
surface of the neutron star. These photons scatter off the incoming
matter; which we take to be hydrogen, and deposit their momentum in
it, by Thomson scattering. At the Eddington limit
\be {\dot M}_{\rm
Edd} = R_6\ 1.5 \times 10^{-8} M_{\rm NS} {\rm yr}^{-1}
\label{eqEdd}
\ee
where $R_6$ is the radius of the neutron star in
units of $10^6$ cm (the latter being a typical neutron star radius)
and $M_{\rm NS}$ is the mass of the neutron star in units of the
mass of our sun $\msun$, the outward pressure from the scattering of
photons by the hydrogen is sufficient to counteract the inward
gravitational force. This, then, gives a maximum rate of accretion
in the neighborhood of rates $\sim {\dot M}_{\rm Edd}$.
${\dot M}_{\rm Edd}$ is the rate of mass accretion which just holds
off the accreting protons,
\be
{\dot M}_{\rm Edd} = 2\times 10^{38} {\rm \ ergs/sec}.
\ee
(We shall
see later on that accretion can begin again at $\dot M \sim 10^4
{\dot M}_{\rm Edd}$, where it is called ``hypercritical accretion.")

Now the processes that we are going to discuss such as common
envelope evolution take a time $\sim$ years. Given the $10^{-8}$ in
Eq.~(\ref{eqEdd}), accretion at the order of the Eddington rate is
unimportant and can be neglected.

\begin{figure}

\centerline{\epsfig{file=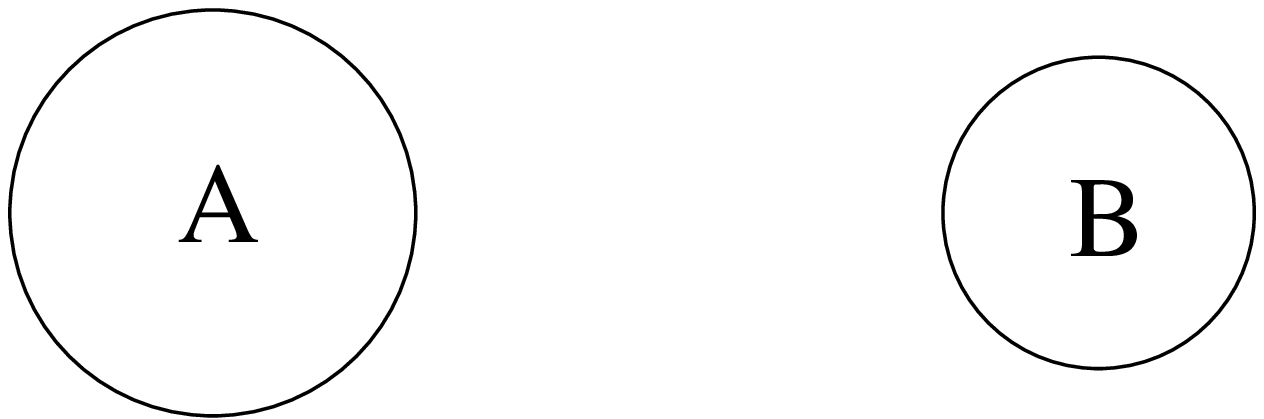,width=3in}} \centerline{$16\msun$
star $\;\;\;\;\;\;\;\;\;\;\;$ $12\msun$ star} \vskip 5mm
\centerline{\epsfig{file=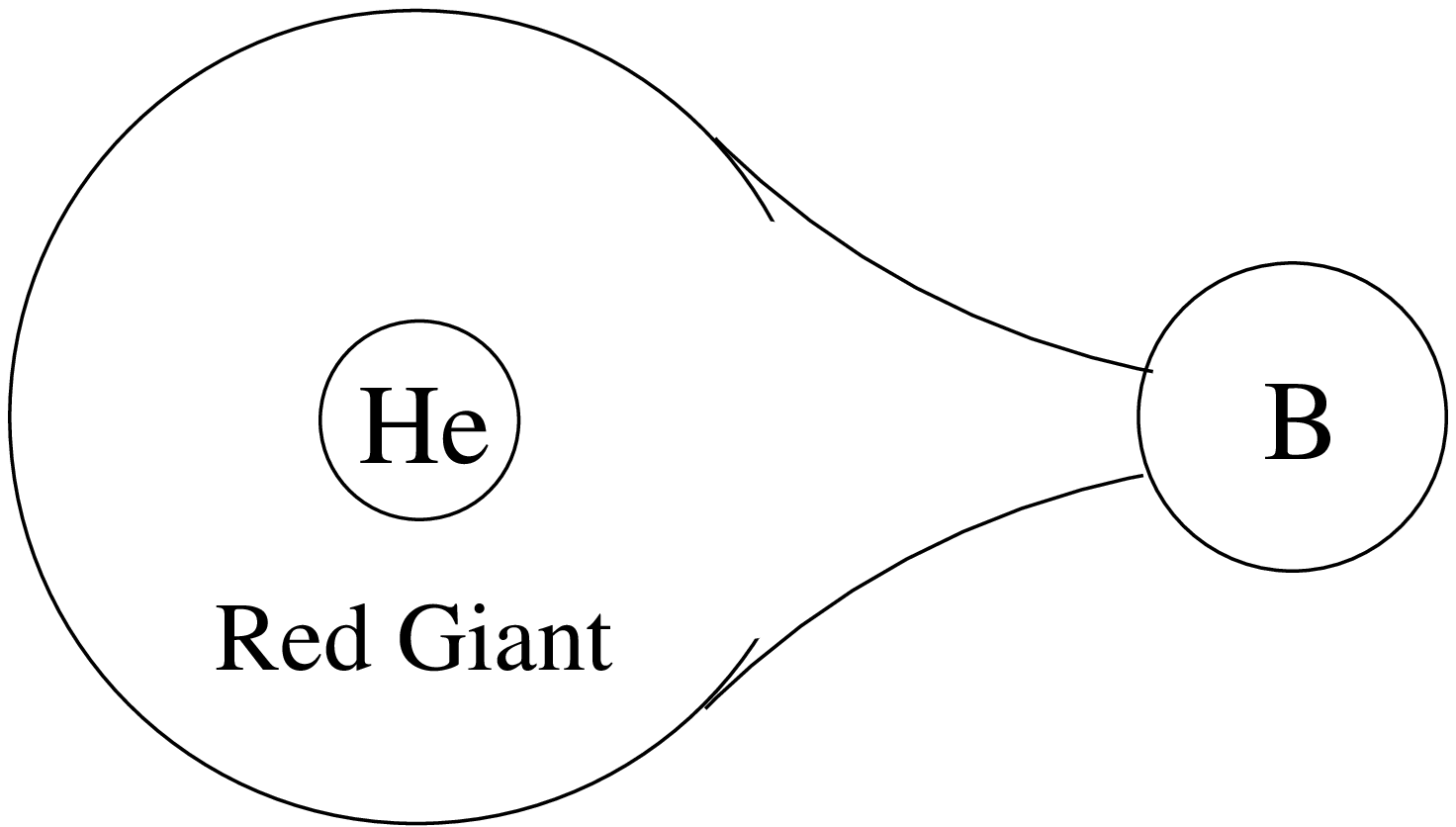,width=3in}} \centerline{Roche
Lobe overflow} \vskip 5mm
\centerline{\epsfig{file=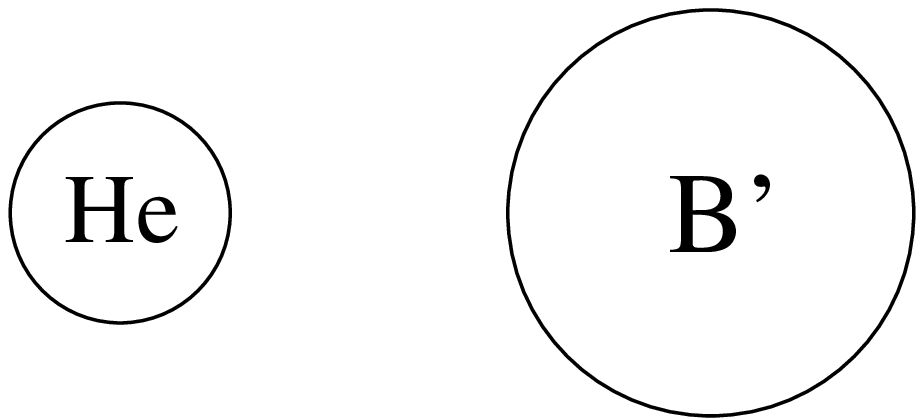,width=3in}} \centerline{He core
$\;\;\;\;\;\;$ $20\msun$ star} \vskip 5mm
\centerline{\epsfig{file=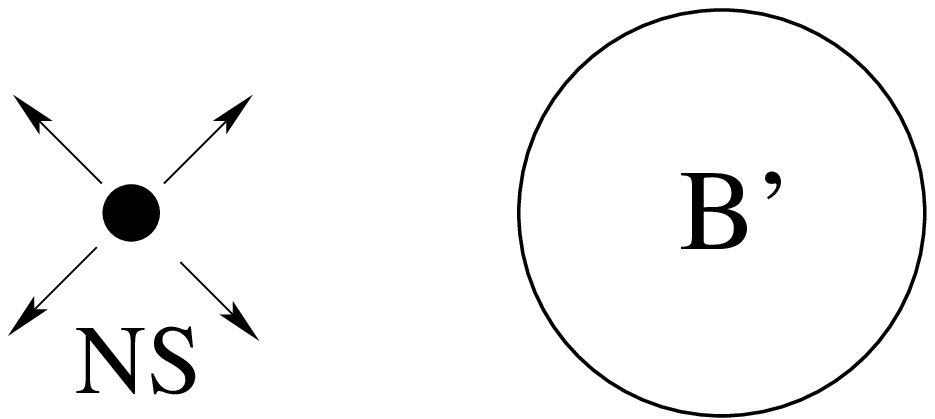,width=3in}} \centerline{Collapse
of He star into a neutron star} \vskip 5mm
\centerline{\epsfig{file=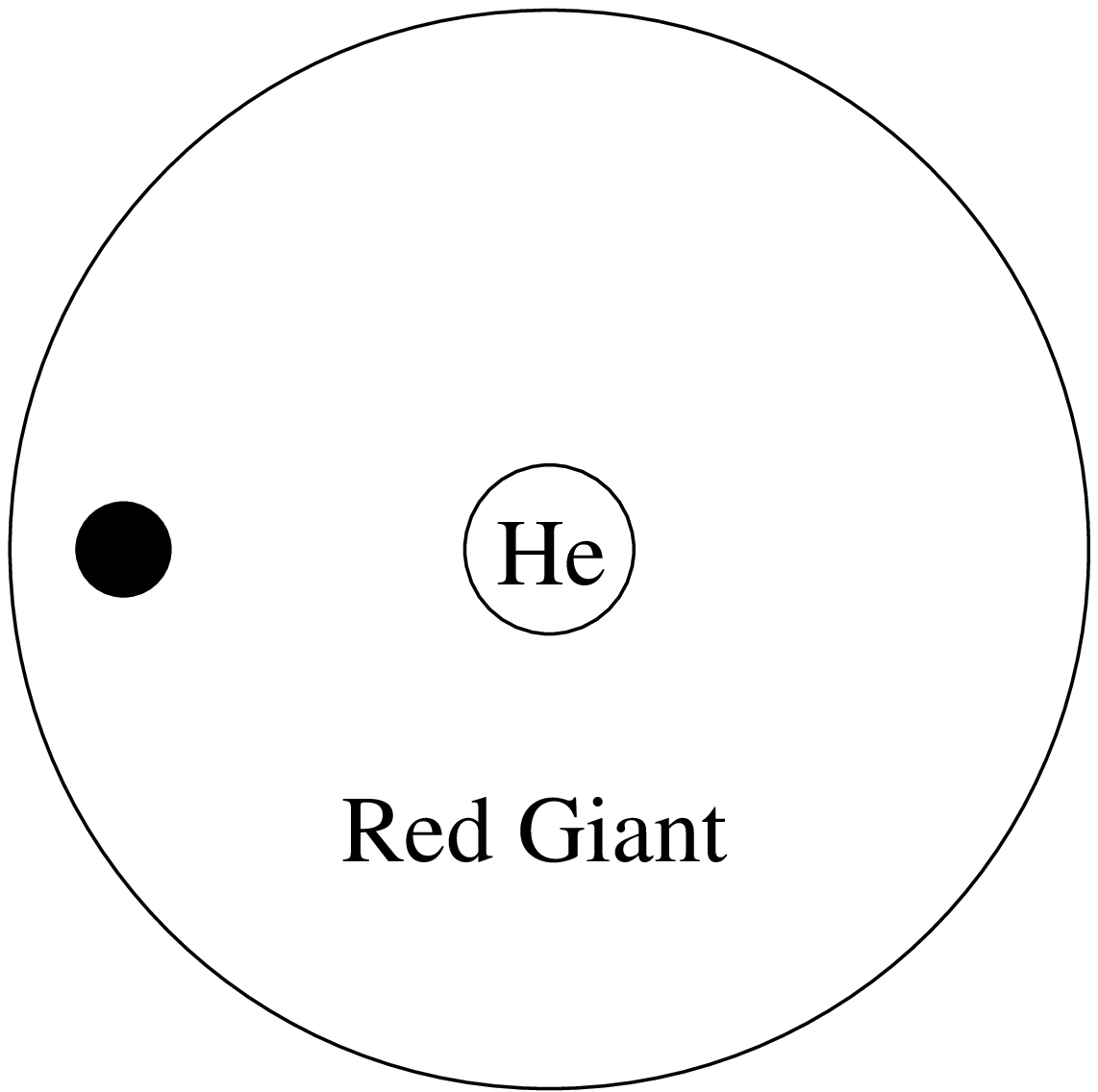,width=3in}} \centerline{$20\msun$
star evolves into red giant, enveloping the neutron star} \vskip 5mm
\caption{Conventional model for double neutron star binary
evolution. The neutron star goes into common envelope evolution,
dropping towards the star, the increased gravitational binding
energy being used to expel the hydrogen envelope. The He star
explodes into a neutron star and if the two neutron stars remain
bound, this is a binary pulsar. }\label{figDNS}
\end{figure}

Let us begin our evolution of binary neutron stars starting from two
giants, with zero age main sequence (ZAMS) masses 16 and 12 $\msun$.
Each star burns hydrogen for $\sim 90\%$ of its lifetime, but the
lifetime of the $16\msun$ star is shorter than that of the $12\msun$.
When the former has finished core burning of hydrogen, it expands in
red giant stage, and the outer part of the red giant crosses the
Roche Lobe, the line of equipotential between the two stars.
The more massive $16\msun$ star evolves first into red giant transferring $\sim 8\msun$ to $12\msun$ star by Roche Lobe overflow leaving the He core and $20\msun$ star as in Fig.~\ref{figDNS}.

In binary evolution usually a flat $q$ ($q=M_2/M_1$) distribution in
the IMF (initial mass function) is assumed. This has the correlation between primary and secondary star in it that the former is more massive than the latter, which must be present because it evolves first. Although offhand this looks like a quite minimal correlation, it produces factor $\sim 2$ difference in results.

We want to zero in on the common envelope evolution of the first
born neutron star with the helium core of the $20\msun$ star. The
common envelope evolution takes only $\sim$ years, so thinking of
Eddington limit for mass accretion of $1.5 \times 10^{-8} \msun$
yr$^{-1}$ the accretion is quite negligible and has generally been
neglected in the literature to date. However it is unreasonable to
assume the Eddington rate to hold forever, as the rate of accretion
continuously increases. Ultimately the density of matter falling onto
the neutron star must be sufficiently high that the photons
originating from the neutron star by random walk in the framework of
the infalling matter actually move out at a lower velocity than the
infalling matter has in falling onto the neutron star, so that the
photons are swept back onto the neutron star by the adiabatic
inflow. Good discussions of this are given by
\cite{Blon86,Brown94b,Houck91,Brown00}.

With the increase in the rate of accretion, which we scale with
Eddington
 \be \dot m = \frac{\dot M}{{\dot M}_{\rm Edd}},
 \ee
in the region of rates $\dot m \sim 1$, the cutting off of increase
in accretion above that value sets in. The accretion from $\dot m
\sim 1$ to $10^4$ may be complicated, the effects from the Eddington
limit causing uneven or sporadic flow, but for $\dot m > 10^4$ where
hypercritical accretion sets in, the accretion again becomes simple
with uniform flow.

The nature may not be so simple because the matter is accreted at
a low temperature, which means that it cannot get rid of its energy
rapidly by neutrino emission, which increases with a high power of
temperature $T$, but is low for the accretion temperatures.

In the scenario of Bethe and Brown~\cite{Beth98}, neutrinos were
trapped in the collapse of large stars, so the trapping of photons
in order to give hypercritical accretion seemed self-evident.
Indeed, one can work out that as the accretion increases to the Eddington
limit, the photon trapping radius moves to the radius of the neutron star $R_{\rm NS} \sim 10^6$ cm. There is, then, a range of $\dot M$'s,
\be
{\dot M}_{\rm Edd} < \dot M < 10^4 {\dot M}_{\rm Edd}
\ee
in which the outcome of accretion is not clear, some matter being accepted by the neutron star and some matter being lost in space. This dees not really concern us because for rates $\sim 10^4 {\dot M}_{\rm Edd}$ and higher, the accreted matter piles up in an accretion shock. As the amount of matter increases, the density at the base of the accretion shock increases and ultimately, when the temperature reaches $T\sim 1$ MeV neutrino pairs  carry off the energy as rapidly as it is accreted, holding the temperature to $\sim 1$ MeV. An analytical calculation of this was carried out by Brown and Weingartner \cite{Brown94b}.

%
%
%
%
Given
the two giant progenitors going through the evolution shown in
Fig.~\ref{figDNS}, and the helium stars, the lowest mass in the
problem was the neutron star mass $M_{\rm NS}$, and the equations
could be solved algebraically if $M_{\rm NS}$ was set equal to zero
\cite{Beth98}. The result was that the first-born neutron star (see
Fig.~\ref{figDNS}) accreted $1\msun$ in the common envelope
evolution during the red giant phase of the remaining giant.
Belczynski et al. \cite{Belc02} (denoted as BKB) removed the
approximation made by Bethe and Brown and obtained the result that
the accretion was $\sim 0.8\msun$ for the lower mass neutron stars
and $\sim 0.9\msun$ for the higher mass ones. In either case it was
sufficient to send the primary into a black hole (see Case D2 of
BKB).
Zeldovich et al. \cite{Zeld72} had found already in 1972 that the
accreting matter piles up, pushing its way up through the accretion
disk. At this stage we go over to the two-dimensional hydro
calculations of Armitage and Livio \cite{Armi00}. They showed that
an accretion disk reformed inside of the accretion shock, allowing
matter to accrete onto the neutron star. They,
however, suggested that jets might drive the hypercritical accreting
matter off, saving the neutron star from going into a black hole.
We find, however, in agreement with observations, that the first born neutron star (the pulsar) is  within 4\% difference from the mass of the companion in the binary,
in agreement with our scenario that hypercritical accretion can be avoided if the two giant progenitors are within 4\% of each other in mass. No sign of any other amount of hypercritical accretion, which could simply add to the mass of the neutron star is seen. If, indeed, the accretion were stopped at some intermediate stage by jets, one would expect a distribution in pulsar masses compared with companion masses that differed from the 4\%. Since no such distribution is seen, our assumption that the pulsar has gone into a black hole (if the giant progenitors differ by more than 4\% in mass) is validated.

Bethe and Brown \cite{Beth98} found that the accretion was $\sim
10^8$ to $10^9$ times Eddington accretion, and assumed that with
this vast amount of matter, one was back to the classical accretion
of Bondi \cite{Bond52} which took no note of the above complication,
but did guide the matter correctly through the sonic point; {\it
i.e.}, enforced the classical requirements for accretion.
The accretion rate could not be determined directly in the Bethe and Brown \cite{Beth98} calculation, but it was compared with the fully three-dimensional numerical calculation of Terman et al. \cite{Term94}

\subsection{Formation Of Double Neutron Star Binaries}

We do see double neutron star binaries, so the first-born neutron
star cannot always go into a black hole. If the two giant progenitors are so close in mass
that they burn helium at the same time, then they would go into a
helium common envelope evolution, the one star sending He across the
Roche Lobe into the other and the other one sending its expanding
helium to the first star~\cite{Brown95}. Helium burning takes up 10\% of the star
lifetime, most of the time being spent in the main sequence hydrogen
burning. To go from lifetimes to masses one must divide by about
2.5, so the two giant progenitors must be within 4\% of each other
in mass.

This was a definite prediction modulo one assumption: That helium is
not accreted during the helium common envelope evolution. The
assumption was justified however by Braun and Langer \cite{Brau95}
who showed that the helium burning time is too short for appreciable
mass to be accepted by either helium star. Thus, the clear
prediction is that two neutron stars in a binary must be within 4\%
of each other in mass.

There is, however, one proviso. That is that the lower mass neutron
stars in the double pulsars J0737$-$3039 and J1756$-$2251 go through
not only the hydrogen burning red giant but also a helium burning
red giant. The temperature for helium burning in low mass stars must
be greater than in higher mass stars in order to ensure a
sufficiently high central temperature because of the energy loss
through the surface. During the helium burning red giant, $\sim 0.1$
to $0.2\msun$  can be deposited on the first born neutron star by
the helium star companion, and the first born neutron star should be
that much more massive than the other, in addition to the possible
$\sim 4\%$ difference in mass because they must burn helium at the
same time.

\begin{table}
\caption{Compilation of NS-NS binaries \cite{Lat07}.
}
\label{tabDNS}
\begin{center}
\begin{tabular}{lllc}
\hline
Object      & Mass ($\msun$) & Companion Mass ($\msun$) & References \\
\hline
J1518$+$49 & 1.56$^{+0.13}_{-0.44}$ & 1.05$^{+0.45}_{-0.11}$ & \cite{Tho99} \\
B1534$+$12 & 1.3332$^{+0.0010}_{-0.0010}$ & 1.3452$^{+0.0010}_{-0.0010}$ & \cite{Tho99} \\
B1913$+$16 & 1.4408$^{+0.0003}_{-0.0003}$ & 1.3873$^{+0.0003}_{-0.0003}$ & \cite{Tho99} \\
B2127$+$11C & 1.349$^{+0.040}_{-0.040}$ & 1.363$^{+0.040}_{-0.040}$ & \cite{Tho99} \\
J0737$-$3039A  & 1.337$^{+0.005}_{-0.005}$
       & 1.250$^{+0.005}_{-0.005}$ (J0737$-$3039B)  & \cite{Lyn04} \\
J1756$-$2251  & 1.40$^{+0.02}_{-0.03}$ & 1.18$^{+0.03}_{-0.02}$ & \cite{Fau04} \\
\hline
\end{tabular}
\end{center}
\end{table}

\begin{figure}[t]
\centerline{\epsfig{file=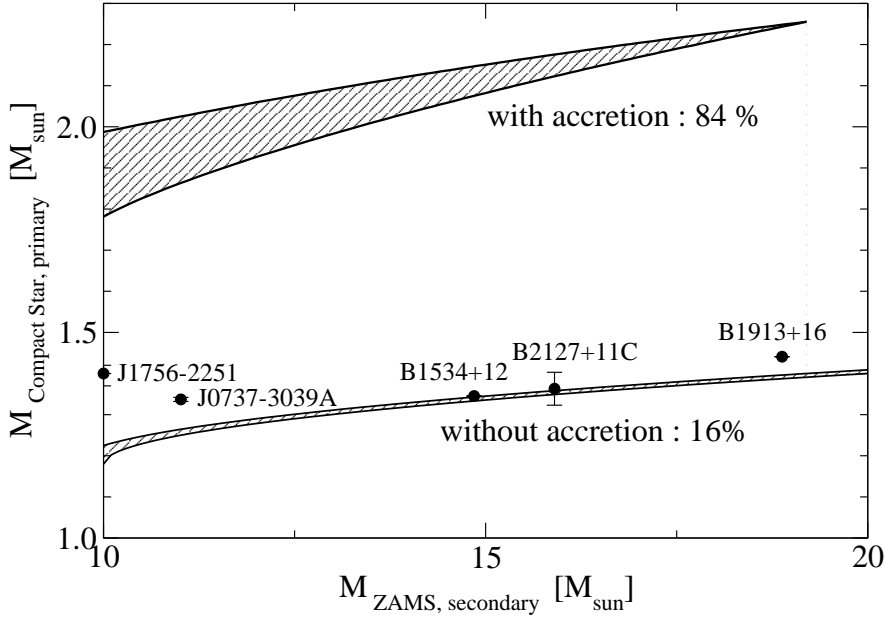,height=4in}} \caption{Masses
of primary compact stars with and without accretion during H red
giant stage of secondary star \cite{Lee07}. The filled circles show
the primary (more massive) pulsar masses in double neutron star
binaries. The corresponding secondary ZAMS masses were obtained from
the secondary (less massive) pulsar masses using eq.~(8) of Lee et
al. \cite{Lee07}. Note that the 84\% corresponds to $M_{\rm
Compact\; Star, primary}> 1.8\msun$. There is uncertainty in the
final primary compact star masses due to the extra mass accretion,
$\sim 0.1 \msun - 0.2\msun$, during He giant stage. This may
increase the primary compact star masses  for both `with-' and
`without-accretion'. Note that with our maximum neutron star mass of
$1.8\msun$, all primary compact stars with accretion would go into
low-mass black holes. }\label{figDNS-f}
\end{figure}

In Table~\ref{tabDNS} we show the compilation of masses of double
neutron star binaries. In J1518$+$49 the uncertainties are too large
to say anything, except that the largest error $+0.45\msun$ is in
the direction of equalizing the masses. In the next three binaries
the masses are very close. In fact, B2127$+$11 is in the globular
cluster and often considered to have resulted from the exchange of
neutron stars formed in different binaries. However, they do satisfy
our nearly equal mass scenario. In the evolutionary calculations of
the double pulsar J0737$-$3039A,B the helium red giant phase has
been put in \cite{Dewi03,Will03}. The $0.1-0.2\msun$ accretion
during the He red giant phase was calculated in hypercritical
accretion by Lee et al. \cite{Lee07}. Using a flat distribution for
the Initial Mass Function, Lee {\it et al}. calculated in
hypercritical accretion what the masses of primary and secondary
neutron stars would be in the scenario shown in Fig.~\ref{figDNS-f}
in which case the primary (first-born) neutron star is assumed not
to go into a black hole, but the neutron star of mass it ends up
with after accretion is assumed to be stable. One sees that the
primary star, the pulsar, has high $\sim 84\%$ probability of ending
up with mass between $1.8$ and $2.3\msun$. This is not seen at all
in Table~\ref{tabDNS}, so we assume that these must all have gone
into black holes.

Note that only the primary pulsar masses (the first born pulsar in
the double pulsar) would be affected by the accretion, the companion
masses lying on the lower thin band in Fig.~\ref{figDNS-f}. In the
figure, we have added the primary pulsar masses of 5 double neutron
star binaries in filled circles. We left out J1518$+$49 because of
the large uncertainty in its masses. We see that the \`a priori
probability for the primary pulsar masses to lie on the lower thin
band within $0.2\msun$ uncertainty due to the possible accretion
during the He giant phase, given no other possibility than to remain
neutron stars, is
\be P= (0.16)^5 \simeq 1\times 10^{-4}.
 \ee
Obviously there is negligible probability assuming random neutron
star masses that the two neutron stars in each of 5 binaries are all
within 4\% of each other in mass.

For our later argument, we note here that the Hulse-Taylor pulsar
B1913$+$16 has mass $1.4408\pm 0.0003 \msun$. It is the most massive
-- and the most accurately measured -- of the neutron stars in
binaries.

\subsection{Neutron Star - White Dwarf Binaries}

In Table~\ref{NSWD}, we list the neutron-star, white-dwarf binaries
in which the central value of the neutron star mass is $\ge
1.5\msun$. We should point out from the neutron star masses in
Table~\ref{NSWD}, given the errors noted, that there is no neutron
star mass other than in J0751$+$1807, which is definitely above
$1.6\msun$. The interval from $1.5 \msun$ to $2.1 \msun$ of
$0.6\msun$ is about half of the total spread, from $1.1\msun$ to
$2.1\msun$ in measured neutron star masses, so it would appear
strange for only one of them to lie in the upper half of the
possible interval.

\begin{table}
\caption{Neutron star masses in neutron-star,
white dwarf binaries in which the central value
of the neutron star mass is $\ge 1.5\msun$ \cite{Lat04}.
}
\label{NSWD}
\begin{center}
\begin{tabular}{cc}
\hline
Name & Pulsar Mass ($\msun$) \\ 
\hline
J1713$+$0747 & $1.54^{+0.007}_{-0.10}$ \\ 
B1855$+$09   & $1.57^{+0.12}_{-0.11}$  \\ 
J0751$+$1807 & $2.10^{+0.20}_{-0.20}$  \\ 
J1804$-$2718 & $<1.70$                 \\ 
J1012$+$5307 & $1.68^{+0.22}_{-0.22}$  \\ 
J0621$+$1002 & $1.70^{+0.12}_{-0.29}$  \\ 
J0437$-$4715 & $1.58^{+0.18}_{-0.18}$  \\ 
J2019$+$2425 & $<1.51$                 \\ 
\hline
\end{tabular}
\end{center}
\end{table}

On the other hand, sufficient of the central values of the neutron
stars in binaries with white dwarf lie above $1.5\msun$ that it
might be prudent to quote the maximum neutron star mass as the Bethe
and Brown \cite{Bethe95} value of $1.56\msun$ deduced from the cut
in the Fe production in 1987A, rather than the earlier $1.5\msun$
\cite{Brown94}.

The conclusion of Lee et al. \cite{Lee07} is that the maximum
neutron star mass must be less than $1.8\msun$, otherwise the
spectrum of double neutron star binaries would be completely
different from that shown in Table~\ref{tabDNS}.

The largest measured mass is in J0751$+$1807, a neutron star in a
binary with white dwarf, with mass $2.1^{+0.20}_{-0.20} \msun$
\cite{Nice05}. However at 95\% confidence level, the mass is
$2.1^{+0.4}_{-0.5} \msun$ so it could be as low as $1.6\msun$ at
this level. Clearly it could be $1.8\msun$ with not unreasonable
probability. As mentioned below, the measured value has recently
been reduced to $1.26^{+0.14}_{-0.12}\msun$.

A number of calculations were made to estimate the mass of the
compact object in SN1987A. They were all more or less in the region
of masses arrived at by Bethe and Brown \cite{Bethe95} from the
$\sim 0.075\msun$ of Ni which was ejected in SN1987A. This gave
an estimate of the separation in the density,
between the mass that fell back onto the compact object and that
which was ejected. The deduced compact object mass was $<1.56\msun$.
Calculations by Thielemann et al. \cite{Thie92} estimated the mass
of the compact object formed in SN1987A to be within the range of
gravitational masses $1.30 - 1.50 \msun$. The higher values follow
from accretion during the formation of the delayed shock. The
delayed scenario is now believed to be the correct one. The
$1.5\msun$ pertaining to SN1987A is supported by the work of Timmes
et al. \cite{Timm90}.

Tauris and Savonije \cite{Taur99,Taur00} evolved neutron star
white-dwarf binaries, mostly by conservative mass transfer, so the
evolution should be reliable. Most of their neutron star masses were
in the neighborhood of $2.1\msun$ whereas all of the other 8
accurately measured neutron star masses in binaries with white
dwarfs \cite{Lat07} are close to $1.5\msun$. There must
be some reason that limits them but we do not know what it is at the moment.

We are well aware that neutron stars in binaries in some globular
clusters have been reported with masses substantially above our
$1.56\msun$ limit. For example, the relativistic periastron advance
for the two eccentric systems in the globular cluster Terzan 5, I
and J, indicates that at least one of these pulsars has a mass of
$1.68\msun$ at 95\% confidence \cite{Rans05}, not so different from
the $1.6\msun$ of the Nice et al. J0751$+$1807 $1.6\msun$ at 95\%
confidence.

The situation in globular clusters is very different from Galactic.
With very low metallicity, their stars resemble more Population II
stars than Galactic ones. This should not make a difference however
in our calculation of kaon condensation and its consequences since
we find the condensation depends only on the baryon number density.

{\it Our statement on kaon condensation is admittedly a very strong
one. But it is a falsifiable prediction and should be
straightforward to prove us wrong: find a well measured double
neutron star binary in which the two neutron stars are more than 4\%
different from each other (modulo some small additional shift by He
red giant) in mass.}

\subsection{Are There Two Branches of Neutron Stars ?}
If the $2.1 \msun$ neutron star of Nice et al. \cite{Nice05} were
firmly established, then there would have to be a loop-hole or an
alternative to the Brown-Bethe scenario for the formation of neutron
stars and light-mass black holes. Indeed Haensel et al.
\cite{Haen07} have proposed an interesting way to reconcile the
Brown and Bethe $M_{\rm NS}^{max}=1.5\msun$ with the Nice et al.
\cite{Nice05} $2.1 \msun$ neutron star, two branches of neutron
stars --- reconciling a $2\msun$ pulsar and SN1987A. Two scenarios
of neutron star formation were considered. Here we interpret the two
scenarios in terms of two possibilities in our language. In the
first scenario the formation of the strangeness core is rapid; this
is the case of the double neutron star binaries in which the
hypercritical accretion from the evolving companion onto the
first-born neutron star, the pulsar, is at a rate of $\sim 10^8
\dot{M}_{\rm Edd}$, so that the total accretion time is $\sim 1$
year. In the second scenario the neutron star is already present
when the main sequence progenitor of the white dwarf begins to
evolve. The neutron star accretes matter (and is spun up) by
accretion from the companion red giant in stable mass transfer. The
rate of mass transfer is $\lsim \dot{M}_{\rm Edd}$, so it is $\sim
10^8$ times slower than in the first case. Models of quark
deconfinement (producing quark matter) are found in which the
condition $M_{\rm NS}^{max} \gsim 2\msun$.

The Haensel et al. reconciliation is ingenious and deserves to be
paid attention. However in view of the recent development on Nice et
al's $2.1 \msun$ neutron star discussed in the next subsection,
reconciliation of that type does not seem called for. We would
nonetheless suggest that the reasoning for the two tracks of Haensel
et al is not valid for the case at hand. Of the two possible phases,
quark matter and kaon condensation,  the former phase would be
favored at high temperature, as is found in collapses like 1987A,
and the latter in a cool scenario like the $\sim 10^8$ yr accretion
from an evolving main sequence star onto the neutron star to make a
neutron-star, white-dwarf binary. In 1987A the collapsing star heats
up, to a temperature $\sim 25$ MeV in the center, because of Joule
heating. The latter is accomplished by the $\sim 100$ MeV highly
degenerate neutrinos leaving their degeneracy energy in the center
of the forming neutron star as they leave with energies determined
by the surface temperature. On the other hand, the temperature from
accretion onto neutron stars in the white-dwarf, neutron-star binary
evolution is much less than 1 MeV, so the thermal neutrino emission
is almost negligible. Thus, the collapse into strangeness
condensation would be favored by the more massive of the two stars
in a double-track scenario and the transition into a black hole by
strangeness condensation in the collapse of a large star would favor
quark matter. In other words, the two tracks of Haensel et al would
be filled in the different way, if indeed they were.
\subsection{New Measurement Of the Neutron Star Mass in J0751+1807}
A new value for the mass of J0751+1807 announced by D. Nice in the
McGill meeting on ``40 Years of Pulsars," August 12-17, 2007,
http://www.ns2007.org, is now $1.26^{+0.14}_{-0.12} \msun$ to
replace the old value $(2.1 \pm 0.2)\msun$. This removes the
necessity for two branches of neutron stars, with which we had
troubles noted just above and reinforces the argument given in
\cite{BBL07} that the 2.1 $\msun$ must be wrong, for were it
correct, the spectrum of neutron-star masses would be completely
different from what it is. Thus, our statement is that {\it all ends
with baryon number density $\sim 3n_0$, there is only
``nothingness," the space-time volume of the black hole, beyond.}
Assuming that we are correct, this leads us to the strong statement
that color superconductivity, color-flavor locking and all of the
other ``exotic" phases that have been suggested for high density
matter lie in the ``black-hole nothingness," being inaccessible by
the observers.

\subsection{Neutron Stars In Globular Clusters}

We have no doubt that neutron stars with masses above our limit of
$M_{NS}(max)=1.56 \msun$ will continue to be reported for globular
clusters.~\footnote{They will surely generate -- as the $2.1\msun$
mass did -- a large number of theoretical publications that will
purport to invalidate the onset of kaon condensation at $\sim 3n_0$
and its crucial role in the fate of compact stars.} No real
evolutions of these have been carried out. From the measurement of
the rate of advance of the angle of periastron of the binaries
combined with matching white dwarfs with the data base, a
Bayesian-type analysis gives the possibilities of different
white-dwarf masses. Together with the mass function, probabilities
of neutron star masses are then obtained.

Binaries with measurable quantities, i.e., those with rapid advances
in the rate of periastron advances, are clearly favored and the
probability of measurability should be built into the Bayesian
analysis.

Why we were so worried about J0751+1807 was that the Shapiro effect
was involved in the analysis. With an accurate Shapiro effect and
mass function, the two masses in the binary can be measured
accurately.
The Shapiro effect is a relativistic effect resulting in the time retardation of radiation from the companion passing by the neutron star. Thus, it can only be measured when the angle of incidence is $\gsim 80^\circ$, so that the neutron star and companion are nearly in the same plane. The measurement of this shift gives a relation between that of the two masses additional to that of the mass function, so with the two relations one can obtain both masses.

Stars in globular clusters have generally low metallicity, more like
Population II and III stars. They tend to be much more massive than
Galactic stars. Our assertion that kaon condensation with subsequent
collapse into black holes depends {\it only on the baryon number
density} will be very interesting to test in the new environment of
globular clusters. Following the present preliminary
``Schadenfreude" of observers trying to prove that theorists are
wrong we look forward to real tests of our assertion in environments
with low metallicity.

\subsection{Convective Carbon Burning And Compact Star Masses}

\begin{figure}
\centerline{\epsfig{file=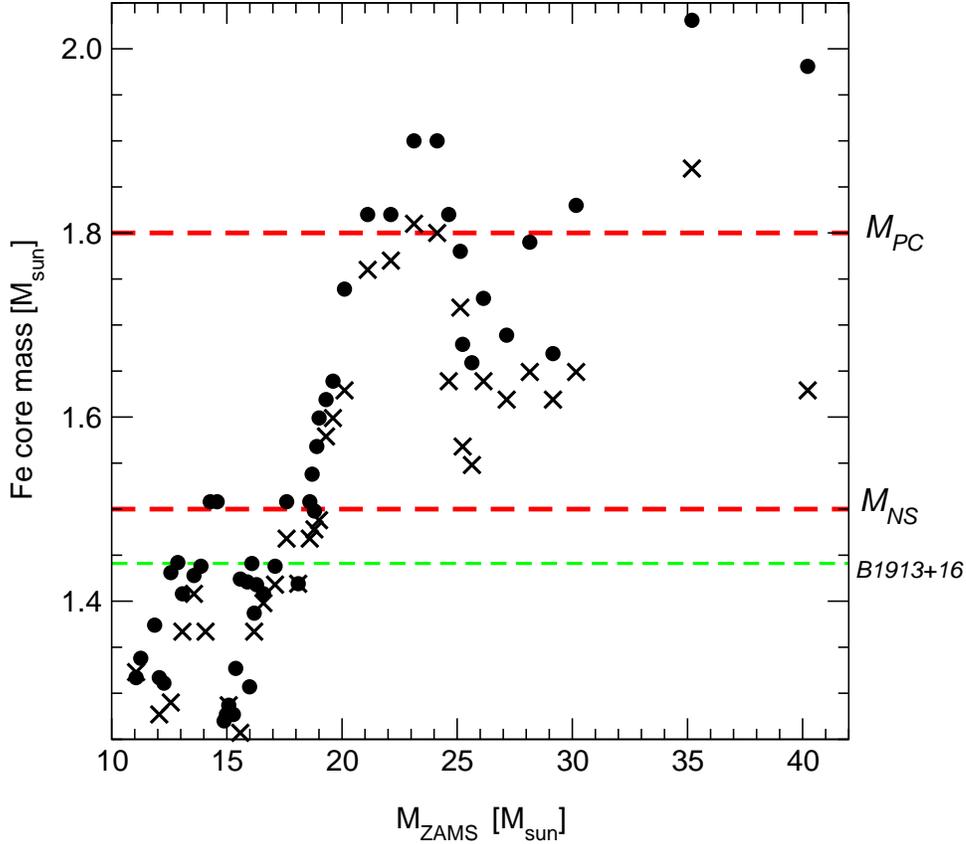,height=14cm}}
\caption{
Comparison of the iron core masses at the time of iron core
implosion for a finely spaced grid of stellar masses \cite{Hege01}.
The circular black dots were calculated with the Woosley \& Weaver
code \cite{Woos95}, whereas the crosses employ the vastly improved
Langanke, Martinez-Pinedo rates \cite{Lang00} for electron capture
and beta decay.
If the assembled core mass is
greater than $M_{\rm PC}= 1.8\msun$, where $M_{PC}$ is the
proto-compact star mass as defined by Brown \& Bethe \cite{Brown94}, there is
no stability and no bounce; the core collapses into a high mass BH.
$M_{NS}=1.5\msun$ denotes the maximum mass of NS \cite{Brown94}.
The mass of the heaviest known well-measured
pulsar, PSR B1913$+$16, is also indicated with dashed horizontal line
\cite{Thor99}.}\label{figFe}
\end{figure}

Although the star in the center of SN1987A disappeared, we do know
that the original star Sanduleak-69$^\circ$202 was in star catalogs
and did have a mass of $\sim 18\msun$. For many years there was
substantial uncertainty in the evolution of giants in the region of
masses in which they would begin, with increasing ZAMS mass, to go
into black holes. The most important nuclear process in this
evolution was the \be ^{12}{\rm C} (\alpha,\gamma) ^{16}{\rm O} \ee
reaction. This reaction determines where the convective burning of
carbon stops. In many years of work, the Stuttgart group
\cite{Kunz01} and more recently Buchmann and Barnes \cite{Buc06} determined this to be
\be
S_{\rm tot}^{300} &=& \left\{ \begin{array}{ll}
                      162\pm 39 {\rm \ keV\ barns} & ({\rm Stuttgart}) \\
                     80^{+20}_{-20} + 53^{+13}_{-18} {\rm \ keV\ barns} & ({\rm Buchmann \ \&\ Barnes})
                     \end{array}. \right.
\ee
The upper superscript 300 means that its
value is given for a reaction energy of 300 keV. The convective
$^{12}$C burning carries off a lot of entropy through neutrino
pairs; it is the first reaction in the stellar burning of stars to
do so. Once the $^{12}$C is sufficiently used up so that this
reaction no longer takes place, the increase in entropy must come
from the increase in size of the Fe core, which has entropy $\sim
3/4$ per nucleon (a bit less than the $\sim$ unity of the original
Bethe et al. paper \cite{BBAL}). We show in Fig.~\ref{figFe} that
the convective $^{12}$C burning ends at $18\msun$ \cite{Brown01b}.
Note that the Fe core mass increases above $1.5\msun$ just at $\sim
18\msun$. (Brown et al. point out that the Fe core mass should be a
good indicator of the compact object mass because the decrease in
mass from the general relativistic effect is compensated for by
fallback in formation of the compact object.) Thus, with the best
available nuclear physics to date, the supernova
progenitor of SN1987A collapsed into a compact object of $\sim
1.5\msun$, not far from the mass constructed by Bethe and Brown from
the $0.075\msun$ of Ni formed in the explosion of
$<1.56\msun$.~\footnote{The standard reply of astronomers to the
statement that if a neutron star were present in the region where
SN1987A exploded, then one should have seen something has been:
``The absence of evidence is not evidence of absence." We do not
believe this to be true because one would have seen something had a
neutron star been there. It could not have just disappeared, leaving
no evidence.}

Note that the low-mass black holes will only be formed from ZAMS
$\sim 18-20\msun$ stars; i.e., black holes of $1.5 - 2.5\msun$. In
fact, the black hole from 1987A is our only example of such a black
hole. The next lowest mass one is the $\sim 5\msun$ black hole in
the transient source GRO~J1655 which comes from an $\sim 30\msun$
giant \cite{Lee02}. We would expect black holes from giants with
ZAMS masses $20-30\msun$ to appear in gamma ray bursts, one of $\sim
3\msun$ from ZAMS mass $\sim 20\msun$ in GRB060218/SN2006aj
\cite{BLW07}.

\subsection{Accretion Onto Neutron Star After Supernova Explosion}

\begin{figure}
\centerline{\epsfig{file=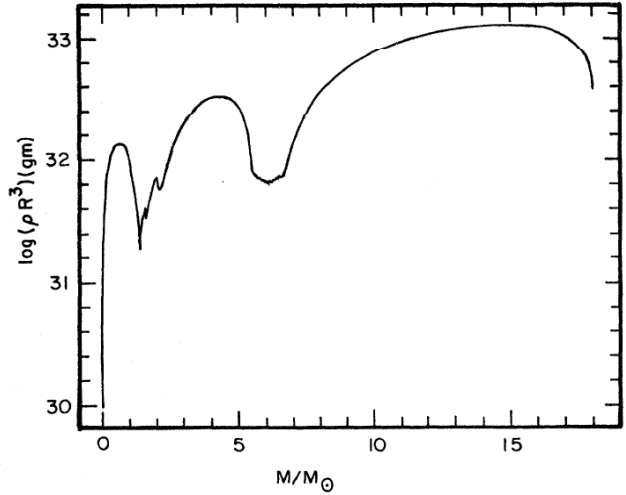,angle=90,height=3in}}
\caption{Distribution of $\rho R^3$ vs included mass in the
presupernova model of a star of mass $18\msun$, according to Woosley
\cite{Bethe90}} \label{figX}
\end{figure}

The shock wave powering the supernova explosion moves outwards
through the helium core, essentially as a Sedov self-similar shock
of constant velocity, the pressure being essentially inversely
proportional to the volume. However $\rho R^3$, where $\rho$ is the
density and $R$ is the radial distance from the center increase
greatly in the hydrogen envelope (see Fig.~\ref{figX}). The shock
has to slow down and the material behind it likewise \cite{Bethe90}.

The shock wave in the supernova after the first four seconds will be
slowed down as it has to pick up more and more material in the
mantle and envelope. The outgoing material in the shock wave thus
has to be decelerated. Within this material, there is a density
discontinuity where the He mantle adjoins the H envelope. In the
deceleration, the lighter H gas has to push the heavier He, the
classical situation for RT instability.

The deceleration takes place because the product $\rho r^3$ increases
going out. The main increase, from $7\times 10^{31}$ to $1.2\times
10^{33}$, takes place continuously in the H envelope, from $m_r = 7
-11 \msun$ (see Fig.~\ref{figX}). In the Sedov theory, a shock wave
continues at constant speed if $\rho r^3=$const because its velocity
is \be U\sim (p/\rho)^{1/2} \ee and $p$ is inversely proportional to
the enclosed volume. But if $\rho r^3$ increases, the shock has to
slow down, and the material behind it likewise. This deceleration
gradually sharpens into an ingoing shock.

Woosley's calculation \cite{Woos90} shows that when the outgoing
shock enters the hydrogen envelope, the deceleration of the shock
sharpens into a reverse shock and this shock has been used to
deposit $\sim 0.1 \msun$ back on the compact object \cite{Chev81}.
Fig.~3 of Shigeyama et al. \cite{Shig88} shows the matter being
deposited back on the neutron star.

When material sent radially inward by the reverse shock reaches the
supernova center, the compact object is no longer there in case the
neutron star receives a kick velocity. The material proceeds to move
outwards from the original center. Eventually all of the matter will
be moving radially outward, through with different speeds depending
on the distance from the center. Thus Chevalier \cite{Chev89}
suggests that the compact object will end up traveling at the same
speed as its kick velocity. Therefore, we may, to a good
approximation, consider the compact object to be at rest with
respect to its ambient matter and use the Bondi \cite{Bond52}
spherical accretion theory to determine $\dot M$, the rate of mass
accretion onto the compact object. Brown and Weingartner
\cite{Brown94b} find that the photons from the initial accretion are
trapped by the high density infalling matter, but that in $\sim 0.6$
yr one would have accretion at basically the Eddington limit.

\subsection{Observability Premium For Unseen Black Hole - Neutron Star Binaries}
Our prediction on the implication of kaon condensation on neutron
stars in a binary has important ramifications on the number of
light-mass black holes. We claim that there should be $\gsim 5$
times more low-mass black-hole, neutron-star binaries than double
neutron-star binaries.

Now what is the status of the observation?

The observed pulsars have all been ``recycled". That is, they have
accreted mass from their companion. Empirically, this accretion of
mass \cite{Taam86} brings down the magnetic field of the pulsar. The
time that a pulsar can be seen is inversely proportional to its
magnetic field. If the magnetic field is brought down a factor of
$\sim 100$ from the recycling, the pulsar will be observable for 100
times longer. It is convenient to introduce the idea of
``observability premium"\cite{Wett96}
 \be \Pi = \frac{10^{12}{\rm G}}{B}
  \ee
where $B$ is the magnetic field of the pulsar. $\Pi\simeq 1$ for
fresh, unrecycled pulsars, which would appear in neutron star, black
hole binaries, because the black hole cannot transfer mass
(``recycle") the neutron star. For PSR~1913$+$16, the Hulse-Taylor
binary, $B\sim 10^{10}$ G for the pulsar, so it will be observable
for $\sim 100$ times longer than the black hole, neutron star
binary. In Table~\ref{tabDNS} we have 6 double neutron star
binaries. Taking 100 to be a typical difference between the
observable times for recycled and unrecycled pulsars, we see that we
should see $\sim 0.6$ neutron-star, black-hole binaries. In fact,
the factor is probably somewhat less than this because the observer
would be somewhat biased against claiming something that cannot be
directly seen, but identifiable only by its effects, such as a black
hole in a binary.

An interesting recent development is the discovery of ``twins"
\cite{Pins06}. These authors show that ``Binaries like to be twins";
i.e., something in their birth process seems to favor binaries in
which the two stars are essentially identical. If this is true, it
would enhance the double neutron star production, because it
increases the number of neutron star pairs that are within 4\% of
each other in mass. They did not, however, correct for selection
effects, which will probably lower somewhat their estimated number
of twins.

Prompted by the paper on twins by Pinsonneault and Stanek
\cite{Pins06}, Lee et al. \cite{Lee07} made an improved calculation
of the ratio of neutron star, low-mass black hole mergers to double
neutron star mergers. The latter could be run up if there were lots
of twins. However selection effects not included in \cite{Pins06}
would have to be taken into account before one could estimate the
number quantitatively. In the non-twin population,  Lee et al. found
that the number of low-mass black-hole, neutron star binaries would
be $\sim 5$ times greater than double neutron-star binaries. For
detection by LIGO the ``chirp" mass would be somewhat greater for
the former than the latter, because of the greater mass for the
black holes.

In summary, it is clear to us that there are $\gsim$ 5 times more
low-mass black hole binaries than double neutron star binaries. As
of now, the merging of these cannot be distinguished in the
short-hard $\gamma$-ray bursts. However the consequences in terms of
the two neutron stars in a binary having to be within 4\% of each
other in mass, with small correction made where there has been some
He red giant mass transfer, are growing in number. We have explained
why, in the case of SN1987A, the absence of evidence is evidence for
a black hole.

\section{CONCLUSIONS}
We have been able to arrive at the strangeness condensation phase
transition critical point $n_c\lsim 3 n_0$ from three different
starting points, the first from the zero-density vacuum, the second
from the vector manifestation fixed point and the third from the
Landau Fermi-liquid fixed point. In the two last cases, the estimate
was made more reliable by arguments based on renormalization group
flow which made all of the quantities whose behavior with density
was unknown rotated out. In particular in the second case, rotated
out were the role of the scalar condensate $\langle\bar q q\rangle$,
the explicit chiral symmetry breaking through $\Sigma_{KN}$, the
role of strange baryons $\Lambda, \Sigma, \Xi$, the $\Lambda(1405)$
which was often viewed as the doorway state for kaon condensation,
all of which figure importantly if one starts from the $T=n=0$
vacuum. We therefore favor the second scenario over the others. (In
\cite{Brown06} a detailed discussion of these matters is given.)

We have shown that the hidden local symmetry in the ``vector
manifestation" (HLS-VM) prescribes the medium dependence imposed by
(Wilsonian) matching to QCD of the quantities entering into the
strangeness condensation. In particular, since HLS-VM is a fixed
point with vanishing gauge coupling constant, chiral perturbation
theory is well-defined. Assuming that HLS theory represents QCD near
the fixed point, one can efficiently calculate the strangeness
condensation by fluctuating about the fixed point. It turns out that
the only degree of freedom we need to take into account is the
vector degree of freedom in the form of on-shell constituent quarks.
In other words, the strangeness condensation can be calculated in
the constituent quark model, with the strange quark a passive
observer.

Although we are at a very early stage of understanding, the recent
development of holographic QCD, in particular, the Sakai-Sugimoto
model, that presumably encapsulates QCD at low energy in terms of an
infinite tower of hidden local fields promises to confirm or
invalidate our main theme. Integrating out the higher members of the
tower leaving the lowest could lead to HLS-VM theory of Harada and
Yamawaki, with the vector manifestation representing the effect of
higher members of the tower when constrained to QCD. At present, it
is not known how to go beyond the large $N_c$ and large 't Hooft
limit in the Sakai-Sugimoto model but once one knows how to handle
certain corrections in the dual sector with appropriate matching to
gauge theory QCD, the characteristic feature of the vector
manifestation such as the hadronic freedom on which our argument
relies could be verified.

In a neutron star the $K^-$ chemical potential $\mu_{K^-}$,
essentially the $K^-$ mass, will be lowered by the attractive
interaction between the quarks in the neutrons and protons, chiefly
by the former, so that at a certain density, which we estimate to be
$3 n_0$, the $\mu_{K^-}$ becomes equal to $\mu_{e^-}$. At that point
the electrons change into $K^-$ mesons, which go into an S-wave Bose
condensate. With the accompanying lowering of pressure, the star
drops into a black hole. This we believe is the first phase transition dense compact star matter undergoes which determines the fate of star, that is whether it becomes a neutron star or a black hole.

We believe that this scenario describes the fate of the supernova
explosion SN1987A. What we are sure of is that in the evolution of
binary neutron stars, the first-born neutron star will go into a
black hole if the two progenitor giants are not within 4\% of each
other in mass. (For the low-mass neutron-star binary the situation
is slightly shifted, as we explained in the text.)

Since our main early interest in kaon condensation was in the
disappearance of evidences of the existence of neutron star in
SN1987A after it had emitted neutrinos for about 12 seconds, we felt
it fitting to adduce even more firm astrophysical observations that
shows that there are a large number of low-mass black holes in the
Universe. Most specific is that we showed that double neutron stars
in binaries must be within 4\% of each other in mass; otherwise the
binary would end up being one of a neutron star and a low-mass black
hole. Observations support the necessity of the neutron star being
within 4\% of each other in mass within any given binary. This also
means that there are $\sim$ 5 times more neutron-star, low-mass
black hole binaries than double neutron-star binaries in the Galaxy.

Finally, putting together all of the above leads us back to the
proposal of Brown and Bethe \cite{Brown94} that the maximum neutron
star mass is $\sim 1.5\msun$, not much higher than the mass of the
Hulse-Taylor pulsar 1913$+$17 of $1.44\msun$. This is the only
accurately measured mass of the more massive neutron stars. We
believe, for example, the probability that J0751$+$1807, of mass
$2.1\pm 0.2$ but with 95\% probability $> 1.6\msun$ has mass $\sim
1.5\msun$ to be substantially higher than for the spectrum of
neutron stars in binaries that would exist were $2.1\msun$ neutron
stars to be stable. The announced reduction of the measured mass of
J0751$+$1807 to $\sim 1.26\msun$ goes in the direction consistent
with the above reasoning.

The Harada and Yamawaki HLS-VM had predictions that countered well
established beliefs. Perhaps the most controversial was the
prediction in HLS-VM of hadronic freedom, that the interactions
between hadrons go to zero as the temperature goes up to $T_\chi$
from below and the density goes up to $n_\chi$ from $n_0$. This
directly contradicted the accepted belief in heavy-ion community
that interactions between hadrons became more plentiful and stronger
as $T$ goes up to $T_c$ from below. We managed to find what we
believe to be the crucial experiment to decide this issue, namely
the STAR experiment of peripheral Au$+$Au collisions which measured
$\rho^0/\pi^-$ ratio \cite{STAR}. We discussed this experiment
briefly in Sec.~5. The experiment was so useful because in this
peripheral experiment the freeze out temperature was equal to the
flash temperature, the temperature at which the hadrons went on
shell. The results of this STAR experiment directly supported the
concept of hadronic freedom in temperature. We are proposing that
the same is the case in density and applies to kaon condensation. We
found in this paper that HLS-VM gave unambiguous statements about
how to handle the medium dependencies and that is why we are so sure
of the results. Needless to say, the crucial element in this
argument is the existence of the vector manifestation (VM) fixed
point which requires the vanishing of the vector meson masses (in
the chiral limit). The VM in Harada-Yamawaki HLS theory is an
unambiguously falsifiable prediction: it could be invalidated if an
unquenched lattice measurement of the spectral function of the
$\rho$ channel $unquestionably$ showed that the (real-time) mass of
the $\rho$ meson does not vanish in the chiral limit at $T=T_\chi$.

The question we have not addressed here is this: if kaon
condensation is inevitable in compact stars once density reaches
$\sim 3n_0$, will other forms of exotic states that can take place
at higher densities, say, $n>n_c \sim 3n_0$, such as color
superconductivity, color flavor locking etc. be relevant for the
physics of compact stars? We can offer no definitive answer to this
question. Our conjecture is that they will be hidden in the
``black-hole nothingness" inaccessible to the observers. The
question can perhaps be answered when holographic QCD becomes
sufficiently sophisticated and predictive.


\section*{Acknowledgments}
G.E.B. was supported in part by the US Department of Energy under
Grant No. DE-FG02-88ER40388.
C.H.L. was supported by Creative Research Initiatives (MEMS Space Telescope) of MOST/KOSEF.



\end{document}